\documentclass[11pt,fleqn]{article}

\usepackage{amsfonts}
\usepackage{amsmath}
\usepackage{amssymb}
\usepackage{mathtools}
\usepackage{mathrsfs}
\usepackage{enumerate}
\usepackage{bbm}
\usepackage{graphicx,epsfig,amsthm,color}
\usepackage{pstricks}
\usepackage{subfig}
\usepackage{graphicx}
\usepackage{units}

\usepackage{caption}
\usepackage{booktabs}
\usepackage{textcomp}
\usepackage{array}
\usepackage{cite}
\usepackage{cancel}

\date{\today}

\newcolumntype{z}[1]{>{\RaggedRight\hspace{0pt}}p{#1}}
\newcolumntype{w}[1]{>{\RaggedRight\hspace{0pt}}p{#1}}
\newcolumntype{v}[1]{>{\Centering\hspace{0pt}}p{#1}}

\def\be{\begin{equation}}
\def\ee{\end{equation}}
\def\bea{\begin{eqnarray}}
\def\eea{\end{eqnarray}}

\def\be{\begin{equation}}
\def\ee{\end{equation}}
\def\bea{\begin{eqnarray}}
\def\eea{\end{eqnarray}}

\def\erp2{{\rm e}^{2\rho}}
\def\erm2{{\rm e}^{-2\rho}}
\def\er4{{\rm e}^{4\rho}}

\def\be{\begin{equation}}
\def\ee{\end{equation}}
\def\bea{\begin{eqnarray}}
\def\eea{\end{eqnarray}}

\def\m0{m_{\nu_{0,i}}}

\def\T0{T_{\nu_0}}

\newcommand{\half}{\frac{1}{2}}

\newcommand{\beqa}{\begin{eqnarray}}
\newcommand{\eeqa}{\end{eqnarray}}
\newcommand{\bpr}{\begin{problem}}
\newcommand{\epr}{\end{problem}}
\newcommand{\bcent}{\begin{center}}
\newcommand{\ecent}{\end{center}}
\newcommand{\bfig}{\begin{figure}}
\newcommand{\efig}{\end{figure}}
\newcommand{\bpc}{\begin{picture}}
\newcommand{\epc}{\end{picture}}

\renewcommand{\and}{A_{0}^{\nu ,D}(s)}

\newcommand{\bee}{\begin{equation}}

\def\beq{\begin{eqnarray}}
\def\eeq{\end{eqnarray}}

\newcommand{\bright}{\begin{flushright}}
\newcommand{\eright}{\end{flushright}}
\newcommand{\bminip}{\begin{minipage}}
\newcommand{\eminip}{\end{minipage}}

\DeclareCaptionLabelSeparator{mysep}{\hspace{3pt}:\hspace{3pt}}
\DeclareCaptionLabelFormat{mypiccap}{Fig.\hspace{3pt}{#2}}
\DeclareCaptionLabelFormat{mytabcap}{Tab.\hspace{3pt}{#2}}

\begin{document}

\date{}
\title{
\vskip 2cm {\bf\huge Chronons in M-theory and F-theory}\\[0.8cm]}

\author{{\sc\normalsize
Andrea Zanzi\!
\!}\\[1cm]
{\normalsize La Gassa 17, CH-7017 Flims, Switzerland}\\
\\
{\normalsize Email: andrea.zanzi@unife.it}\\[1cm]
}
\maketitle \thispagestyle{empty}

\begin{abstract}
{We consider a model originally proposed in the framework of supergravity in singular spaces. This model is considered as a good summary of the expected features of M-theory and F-theory. The chronon is identified as the fundamental degree of freedom of Nature. The physics of time is discussed in some detail. Some of the possible phenomenological consequences of this scenario are analyzed.}
\end{abstract}
\clearpage

\newpage \thispagestyle{empty} \begin{minipage}[t]{14.4 cm}
\hspace*{-6.5em}

\mbox{}\\[9.4em]
\hspace*{-15.1em}
\mbox{}\\[4.1em]

{\hfill This article is respectfully dedicated}

{\hfill to the memory of Mario Tonin}

 \end{minipage}
\mbox{}\\[-.0 em]
\clearpage

\tableofcontents
\newpage

\setcounter{equation}{0}
\section{Introduction}

One of the greatest open problems in theoretical physics is the quantization of gravity. A detailed quantum description of electroweak and strong interactions is provided by the Standard Model (SM) of particle physics. However, the standard theory of gravity is still Einstein's General Relativity (GR), namely a {\it classical} theory formulated roughly 100 years ago (see e.g. the excellent GR book \cite{Straumann:2013ns}). In order to quantize gravity, a number of proposals have been discussed (see e.g. \cite{Hamber:2009zz, Kiefer:2007ria, Rovelli:2004tv, Blumenhagen:2013fgp}). In this article we will start considering M-theory (for an introduction the reader is referred to \cite{Becker:2007zj}). 

M-theory is a theory of extended objects and it is formulated in 11 dimensions (11D). It can be obtained, for example, considering weakly coupled heterotic string theory in 10D and then moving to the strong coupling region. In this way we are led to heterotic-M-theory, the 11th dimension appears and, at intermediate energies, the theory admits a 5D description. For an introduction to heterotic-M-theory the reader is referred to  \cite{Ovrut:2002hi, Krause:1900zz, Waldram:2001my} and references therein. Let us briefly summarize the resulting scenario. The 5th dimension is spacelike, it is compactified on a $S^1/\mathbb{Z}_2$ orbifold and, hence, two orbifold fixed points are present. In these fixed points we find two 3-branes (one for each fixed point) hosting $E_8$ gauge groups (one for each brane) and the remaining 6 spacelike extradimensions are assumed to be compactified on a manifold typically characterized by a much smaller volume. It is possible to include a larger number of 3-branes obtained from M5-branes compactified on 2-cycles. This multi-brane set up will be the starting point of our paper. 

In \cite{Zanzi:2016thx}, a multi-brane scenario has been exploited to obtain a specific 4D lagrangian containing a scale invariant sector (originally considered by Fujii in \cite{Fujii:2002sb}) and a symmetry breaking sector (i.e. a scale non-invariant sector).  Therefore, the resulting model has been called Modified Fujii Model (MFM). Our braneworld scenario has been originally presented in the framework of supergravity in singular spaces in \cite{Bergshoeff:2000zn} and it will be considered here as a dependable summary of the expected features of M-theory. The MFM lagrangian has been useful to keep under control the cosmological constant \cite{Zanzi:2010rs}, to discuss the collapse of the wave function \cite{Zanzi:2015evj}, to analyze certain aspects of solar physics \cite{Zanzi:2014aia}, to obtain a modified Schroedinger equation \cite{Zanzi:2016yrg}, to formulate an equivalence principle for quantum gravity \cite{Zanzi:2015evj}, to introduce chameleonic matter \cite{Zanzi:2012ha}. 

Typically, a model obtained from string/M-theory should satisfy a certain number of requirements in order to be phenomenologically interesting. Standard problems of string phenomenology and model building are 1) the vacuum energy should be under control, 2) moduli stabilization should be discussed, 3) a reasonable description of supersymmetry (SUSY) breaking should be obtained. Our model is no exception. Let us discuss these three issues separately. This will be a useful way to summarize the results of our paper.

{\bf Problem 1: the cosmological constant} 

The vacuum energy problem is really acute only when we move to the quantum gravity (QG) regime. Indeed, as long as we work with a classical theory of gravity, we can choose the related classical cosmological constant in the way we prefer and this gives us the chance to organize a fine-tuning between the classical cosmological constant of the gravity sector and the quantum contributions to the vacuum energy coming from the remaining interactions. Needless to say, a fine-tuning is not welcome, but at least it can be performed. In QG, on the other hand, this freedom is lost. In this article, the cosmological constant will be kept under control by scale symmetry (and global SUSY). In particular, following \cite{Zanzi:2010rs}, the amount of scale symmetry in 4D will be related to the matter density of the environment. This result can be obtained starting from the 5D braneworld model mentioned above and splitting the branes into two groups: a stack of N 3-branes whose position in the bulk will be parametrized by one modulus and also another (so called ''hidden'') 3-brane whose position in the bulk will be parametrized by another modulus. Hence, we are led to a 4D bi-scalar-tensor effective action. These two degrees of freedom (d.o.f.) parametrize the position of the center of mass of the branes (the dilaton) and the separation between the branes (i.e. the distance between the hidden brane and the stack of branes - this distance will be parametrized by the radion). Dilaton and radion are chameleon fields in our model. This means that their mass is an increasing function of the matter density of the environment and a coupling to baryons is present: these two scalar d.o.f. are locally massive but globally massless. For an introduction to chameleon fields the reader is referred to \cite{Zanzi:2015cch} and references therein, while the original papers about the chameleon mechanism are \cite{Khoury:2003aq, Khoury:2003rn}. As already discussed in \cite{Zanzi:2010rs}, in our model the amount of 4D scale symmetry is parametrized by the dilaton. This means that, locally, scale symmetry is broken and the vacuum energy is large, while globally (i.e. on cosmological distances, where the matter density is small) scale invariance is restored and the cosmological constant is under control.

The reader might say that this does not solve the cosmological constant problem because, in the effective action of \cite{Zanzi:2010rs}, gravity is treated at a semiclassical level at best, but a quantum treatment is necessary. How can we quantize gravity on the brane in this model? In \cite{Zanzi:2015evj}, a Chameleonic Equivalence Principle (CEP) has been discussed. This principle tells us that quantum gravitation is equivalent to a conformal anomaly. Indeed, in our model, whatever will be the source of 4D gravitational field we consider, the chameleonic dilaton will be coupled to it and this creates a shift in the value of the field. A chameleonic jump from one ground state to another one is obtained and this jump is summarized by a conformal anomaly which gives a mass to the dilaton field. Remarkably, a conformal anomaly is a quantum phenomenon and, therefore, the CEP is an equivalence principle for QG. Therefore, the next question is: what are the relevant d.o.f. in QG? This leads us to the issue of moduli stabilization.

{\bf Problem 2: moduli stabilization and dynamics}\\
{\it Exact stabilizing potential}:\\
 The chameleon mechanism is one of the possible ways to give a mass to a scalar field and, therefore, it is a stabilization mechanism. The chameleonic behaviour of the radion has been discussed in \cite{Brax:2004ym}. In order to obtain a chameleonic behaviour of the dilaton in this model, we must fix the value of the dilaton in the UV region/high redshift Universe (this point has been clarified in \cite{Zanzi:2010rs, Zanzi:2012du}). This dilaton stabilization problem is particularly challenging because, in our model, the high redshift Universe corresponds to the strong coupling region of the theory and, consequently, the stabilizing potential must be non-approximated. How can we obtain an exact stabilizing potential for the dilaton in this model? As we will see, the solution to this problem is provided by a 5D black hole localized in one of the two orbifold fixed points (the black hole in the remaining orbifold fixed point is simply the S-dual of the first black hole, hence the two black holes can be identified). Let us discuss this issue in more detail. Typically this curvature singularity is presented as a naked singularity. An horizon for this singularity would be welcome. In order to make some progress, we can map the string frame (S-frame) 5D description of the model into a 5D ball following the prescription of \cite{Zanzi:2006xr}: exploiting a local conformal transformation the singularity is mapped into the bag center, the hidden brane is mapped into an $S^4$ internal boundary, the orbifolded fifth dimension is mapped into a radial coordinate of the bag, the stack of N 3-branes is mapped into an outer $S^4$ boundary. We obtained a 5D potential well with two $S^4$ spaces as boundaries. However, one problem must be faced: in our model (in 5D and in 4D) we do not have local conformal symmetry and, therefore, the 5D bag description seems to be only a useful set-up to perform explicit calculations, but we are forced to come back to the S-frame description (see also reference \cite{Zanzi:2012bf}). However, this bag description will turn out to be much more physical than it seems to be at first sight. This point requires a careful discussion, but it is better to analyze the various issues stepwise. Therefore, let us proceed with the problem of the exact stabilizing potential for the dilaton. If we consider the 5D bag description as physical, this potential well is supporting the possibility that the $S^4$ boundaries are horizons for the black hole. To be more precise, since in 4D the horizon is $S^2$, then in 5D the (spherical) horizon is $S^3$ (not $S^4$) and this is the first clue of a second timelike dimension in the model (we will discuss this point over and over again in this paper). It is common knowledge that, when a horizon is crossed (e.g. we can consider the Schwarzschild black hole as a useful example), then, inside the horizon, the radial coordinate becomes timelike. This is a simple way to obtain an orbifold of time. We are led to an orbifold of time along the bag radius. This peculiar compactification of time has been discussed in \cite{Zanzi:2016thx} (in that paper the idea was to exchange the fifth coordinate with time granted that an Euclidean metric was present, however, here we presented the argument related to the horizon crossing because it seems to be more intuitive and even more general). This picture is taking shape: our universe is located on a $S^4$ manifold and the outer horizon of the 5D black hole, which is an $S^3$ space, is part of this $S^4$ manifold.

Now to the point. The radial coordinate of the bag is timelike and this is the key to stabilize the dilaton in strong coupling with a non-approximated potential. Indeed, let us consider a small quantum fluctuation near the black hole. In this case the theory is strongly coupled and, in harmony with the Horava-Witten analysis \cite{Horava:1995qa, Horava:1996ma}, the size of the fifth dimension is characterized by a particle physics scale, namely it is microscopic. Therefore, these quantum fluctuations are forced to remain inside this small orbifold. However, the relativity of time allows the presence of an asymptotic observer, who perceives the orbifold of time as extremely large and, for this observer, the quantum fluctuations are stretched to macroscopic scales. This is an example of inflation, {\it a relativity-induced inflation}. We infer that, in our model, the dilaton is stabilized in strong coupling in a non-approximated way, simply because the value and the fluctuations of the dilaton are forced to remain inside the orbifold of time and this orbifold is small in strong coupling: the dilaton is stabilized from the standpoint of an asymptotic observer and this is precisely what we need to make the dilaton a chameleon field. \\
{\it The chronon and the bag}:\\
 Now that we touched upon the exact stabilizing potential for the dilaton, we can come back to the two problems mentioned above: 1) what are the relevant d.o.f. in QG? 2) Why the bag description is more physical than it seems to be at first sight? To answer these questions we can use the CEP. The selection of a chameleonic ground state for the dilaton is related to the gravitational field and it is completely analogous to a gauge fixing procedure in a local gauge theory. Following this insight, we will show in this article that the correct group to describe QG is a 4D local conformal group. The gauge fixing procedure is the origin of the chameleon mechanism\footnote{Needless to say, the reader is assumed to be familiar with the chameleon mechanism (for a review, see \cite{Zanzi:2015cch}).}: the chameleon is coupled to the SM fields and this is completely analogous to the relation present in Quantum ElectroDynamics (QED) between the gauge parameter and the non-gauge-fixed fields. However, local gauge symmetries are related to a redundant description of the theory and, therefore, they can be used to remove redundancies (fixing the gauge). The 4D local conformal group is our gauge group for QG and we can remove the redundant d.o.f. coupled to the chameleons: in this way, as we will see, the SM becomes redundant in the local ground state where we quantize gravity and the only relevant d.o.f. for QG is the chronon (a quantum of time) with its two representative fields which are the dilaton and the radion. Indeed, these two chameleon fields play the role of two timelike coordinates (i.e. the quanta of the fields form a string that we interpret macroscopically as a coordinate) and the chameleonic behaviour of the fields summarizes the relativity of time in the gravitational field of the 5D black hole. 

Now we move to the second point: why the bag description is physical? In the 4D MFM lagrangian we have global scale invariance in the IR. However, as already discussed in \cite{Dorigoni:2009ra}, a scale invariant theory is expected to be also conformal when unitarity is not violated. Therefore, we safely assume that the MFM is locally conformal invariant in the IR. Exploiting S-duality, conformal symmetry is present not only in the IR, but also in the UV. 

This is a good place to mention another problem, namely the apparent absence of 5D conformal symmetry. Indeed, the conformal transformation written in \cite{Zanzi:2006xr} is 5-dimensional but we have only 4D conformal invariance. As we will see, the theory is dual to another theory where the observer is located at the center of the bag. In this description the radial coordinate is {\it spacelike} and the 3D universe we observe is simply the bag with its two times. Happily one of the two times is absent in the IR and, hence, the 4D conformal invariance is exactly the symmetry needed to obtain the IR 5D bag. The dual 5D description has two times on the brane, two spacelike coordinates on the brane and one spacelike coordinate in the bulk. We will infer that the bag description is trustworthy because 4D conformal invariance is enough to build the bag and it tells us the geometry of time. For further details about two-time physics see e.g. \cite{Araya:2013bca, Bars:2010zz} and references therein. Enough with the dual model. Let us come back to our model with the orbifold of time.\\
{\it Cyclic cosmology:}\\
The orbifold of time is a useful tool to obtain a cyclic cosmological model. Indeed, before we impose the $\mathbb{Z}_2$ parity of the orbifold, we have an $S^1$ manifold of dilatonic time. The upper half circle ($t<0$) is a Pre-Big-Bang phase (for PBB physics the reader is referred to \cite{Gasperini:libro} and related references), the lower half circle ($t>0$) is the post big bang phase. The cosmological post big bang expansion is mapped into a cosmological PBB contraction by the $\mathbb{Z}_2$ parity. We obtained a bouncing cosmological model. Remarkably, near the big bang, the radion becomes dynamical and the final outcome will be that the various cosmological cycles describe one dilatonic time dimension for each cycle and we obtain a ''flower of time'': the petals of the flower are dilatonic time dimensions described by the chronon during the past cycles of the cosmological evolution. What happens if we modify the matter density of the environment? The value of the dilaton is modified by the chameleon mechanism, but also the value of the chameleonic radion will be (slightly, because the radion is stabilized by some UV dynamics) modified. The shift in the radion selects a totally different dilatonic time dimension. As we will see, this mechanism is supporting the possibility of receiving gravitational signals from the future of a different dilatonic time dimension where the entire universe is almost identical to our universe and where the cosmological evolution has already taken place.\\

{\it The magnifying lens}:\\
In this model the relativity-induced inflation stretches Quantum Mechanics (QM) to macroscopic scales. In particular, as we will see, the enhancement factor is an exponential function of the dilaton. Therefore, when the matter density is small, quantum fluctuations are stretched on very large distances. Consequently, we have quantum physics in the IR. The reader might disagree because classical physics is obviously valid in the Solar System and, after all, in the Solar System the matter density is very small. The answer at this stage is that, in our model, the exponential ''magnifying lens'' is so effective that on Solar System scales we do not perceive the fluctuations anymore. Indeed, when the wavelength of a quantum fluctuation is much larger than the size of the Solar System, classical physics is a good approximation.

{\bf Problem 3: SUSY breaking and the SM}

As we will see, in our model, QG on the brane shows effects which are reminiscent of Newtonian gravity. The presence of scale invariance in the far IR forbids a gravitino mass and, therefore, we have a supersymmetric theory on cosmological distances. Our IR theory is described by global SUSY together with a sort of Newtonian theory of gravity. However, locally, where the matter density is large, the vacuum energy is positive and SUSY is broken. Since the theory has this peculiar Newtonian character on the brane, the Lorentz group does not play a fundamental role anymore and, hence, we obtain a scenario which is (surprisingly!) spin-independent and also, for other reasons that we will discuss, representation-independent. Consequently, the scalar field representative of matter fields which is present in the MFM lagrangian is not less physical than a fermion and, moreover, it is representative of the entire chiral superfield where matter is inserted. These ideas will lead us to a new scenario for SUSY phenomenology: the SM is obtained directly from the chronon (which is a gauge singlet) and SM fields are representative of their entire superfields. Consequently, the recently discovered Higgs boson at LHC is, in our model, a chronon and low-energy SUSY is not absent in our experiments, it is simply represented by SM particles in harmony with our spin (and representation) independence.  \\
These issues concerning SUSY breaking/SM and chronons will be a crucial element of our discussion. For this reason, it seems worthwhile to summarize very briefly the entire ''timelike structure'' that we will obtain. As we will see, a crucial element of our analysis will be the presence of more than one timelike dimension in the model.  Indeed, in this model, we have 2 times in M-theory and 3 times in F-theory. The reader might be puzzled by the presence of multiple times. However, this possibility has already been discussed in the literature (see e.g. the articles \cite{Bars:1997bz, Bars:2006dy}, the book \cite{Bars:2010zz} and references therein). If we consider the standard everyday time, which, in this model, corresponds to the dilaton, the peculiar orbifold compactification will give us a scenario (see also \cite{Zanzi:2016thx}) with a Pre-Big-Bang phase (negative dilatonic time) where the universe is contracting and a post-big-bang phase (positive dilatonic time, where the universe is expanding). The radion is a second timelike dimension and it corresponds to an angular variable, while the 12-th dimension of F-theory will be a 3rd timelike dimension.  \\

In the final part of the paper we will consider more phenomenological issues: 1) the formation of planetary systems; 2) the data of VIRGO. The main idea regarding planets is that, in our model, the exponential ''magnifying lens'' is so effective that on Solar System scales we do not perceive the fluctuations anymore: their wavelength is much larger than the size of the Solar System and, hence, classical physics is a good approximation. However, physics remains ''secretely'' quantum and, therefore, one might search for some quantum signature of our model in planetary systems. Where should we search? Classical physics is good for the Solar System {\it today}. However, 5 Gy ago, during the formation of the Solar System, the dilatonic time was much smaller and, therefore, the dilatonic magnifying lens was less effective than today: we are led to a quantum formation of planetary systems. We will discuss this point in the article.\\

{\bf Summary:}\\

It seems worthwhile to summarize the main original results of this article:
\begin{itemize}
\item Gravity is dynamical also in the IR region even if we switch off the background value of the Planck ''mass''. Indeed quantum fluctuations are enhanced in the IR and, therefore, in this region we obtain a non-minimal coupling term. Happily this term does not clash with the restoration of scale invariance suggested in \cite{Zanzi:2010rs}. 
\item In strong coupling we have a Horava-Witten model and, hence, quantum fluctuations in the 5th dimension take place on scales which are much smaller than cosmological scales. However, the relativity of time allows the presence of an asymptotic observer and in this case cosmological distances are present. In order to analyze the dynamical evolution of the dilaton-chronon for the asymptotic observer (i.e. the cosmological cyclic expansion/contraction) we do not exploit a relativistic formalism in this paper (the idea behind this choice is that whenever the theory is in a high-energy relativistic regime, string dualities can map the theory into a low-energy non-relativistic regime). Instead, we will work with non relativistic quantum mechanics (the need for a quantum description is in harmony with the inflation of quantum fluctuations during the cosmic expansion) and in particular we will exploit coherent states to mimic the oscillation of the chronon along the orbifold of time. 
\item The dilaton is stabilized in strong coupling from the standpoint of the asymptotic observer and this result is non-approximated.
\item A new uplifting mechanism for the vacuum energy has been suggested. We called it {\it orbifold induced uplifting}.
\item The meaning of time operator is discussed.
\item The M-theory model has two timelike dimensions, namely the dilaton and the radion. We can formulate our effective theory on a $T^2$ torus of time and, remarkably, this construction will turn out to be holographic in the sense of \cite{Cohen:1998zx}. Hence, our model is an example of holographic Dark Energy (for a recent review see \cite{Wang:2016och}) and this will lead us to many interesting insights about the high redshift description of our model. 
\item As we will see, the observer cannot analyze quantum gravity in his ground state (i.e. the observer cannot analyze quantum gravity with an experiment performed in his laboratory), however, he can analyze quantum gravity in a different ground state (e.g. in cosmology). The final outcome is that the chronons provide a UV completion of string theory and they are the fundamental d.o.f. of Nature. Since the dynamics of time is governed by a modified Schroedinger equation, we infer that, at this stage, in our scenario, the Schroedinger equation is the most fundamental equation of Physics (see also \cite{Preparata:1997ap} for a similar approach). 
\item It seems at this stage that this model supports the possibility of receiving gravitational waves from the future (of a different dilatonic time dimension).
\item As we will see, in the IR region new gravitational phenomena are expected when we compactify (the dilatonic) time on the orbifold. These new effects are intrinsically extradimensional, it remains to be seen whether they can be described in terms of a field theory on the brane and, in this sense, they look like an example of a Newtonian action-at-a-distance when they are studied on the brane. 
\item The vacuum energy is under control and it is related to the Large Number Hypothesis of Dirac.
\item The model is supporting a surprising spin and representation independence. 
\item The relativity of time is induced by S-duality.
\item In the final part of our paper we will work with a 3-dimensional time in F-theory and we will build 3-dimensional structures made of timelike lines. The geometry of this timelike architecture is reminiscent of the Library of Borges. 
\item The role of space will be discussed and this will lead us to a 6D Library. The final effective theory will be a 4D theory with euclidean time.
\item In our model the formation of planetary-satellite systems is a quantum process. The regions where planets (or satellites) are found correspond to regions of constructive quantum interference. 
\item The role of duality and its connection to the recent detection of GW by VIRGO will be discussed. 
\end{itemize}

A word of caution is necessary about two issues: 1) the way we use string dualities in this model must be further analyzed; 2) we assume that the model becomes classical in the deep UV region. Further research efforts are necessary to make these issues clear.

As far as the organization of this article is concerned, in section 2 the model is presented; section 3 gathers some useful elements from the literature (no original results are discussed). Sections 4-6 contain the original results of this article. Some concluding remarks are summarized in section 7.

\setcounter{equation}{0}
\section{The M-theory model}
\label{model}

This paragraph contains the lagrangian of our model. We called the 4D model Modified Fujii's Model (MFM) because one sector of our lagrangian had already been considered by Fujii (see for example \cite{Fujii:2003pa, Fujii:2002sb}).

We write the string frame lagrangian in the strong coupling of heterotic theory as (the gauge part is not written explicitly but it is present in the theory)
\beq {\cal L}={\cal L}_{SI} + {\cal L}_{SB}, \label{Ltotale}\eeq where the
scale-invariant (SI) part of the Lagrangian is given by (see also \cite{Zanzi:2010rs}):

\begin{equation}
{\cal L}_{\rm SI}=\sqrt{-g}\left( \half \xi\phi^2 R -
    \half\epsilon g^{\mu\nu}\partial_{\mu}\phi\partial_{\nu}\phi -\half g^{\mu\nu}\partial_\mu\Phi \partial_\nu\Phi
    - \frac{1}{4} f \phi^2\Phi^2 - \frac{\lambda_{\Phi}}{4!} \Phi^4 - \frac{\lambda_{\phi}}{4!} \phi^4
    \right).
\label{bsl1-96}
\end{equation}
$\Phi$ is a scalar field representative of matter fields,
$\epsilon=-1$, $\left( 6+\epsilon\xi^{-1} \right)\equiv
\zeta^{-2}\simeq 1$, $f<0$ and $\lambda_{\Phi}>0$.
One may write also terms like $\phi^3 \Phi$, $\phi \Phi^3$  which are multiplied by dimensionless couplings. However
we will not include these terms in the lagrangian. 

The symmetry breaking Lagrangian
${\cal L_{SB}}$ is a stabilizing potential for the dilaton.

In 5D there is a bulk dilaton $C$.
If we make the ansatz 
\bea
ds^2=dz^2+a(z)^2 g_{\mu\nu}dx^\mu dx^\nu
\eea

for the line element in 5D and if we use an exponential superpotential of the form ($k$ is constant)
\bea
U_B=4ke^{\alpha C}, \alpha \in \mathbb{R}
\eea
as suggested by supergravity (SUGRA), then the bulk equations can be written in a BPS form and solved. The solution for the dilaton is
\bea
C(z)=-\frac{1}{\alpha}ln(1-4k \alpha^2 z)
\eea
while for the scale factor is
\bea
a(z)=(1-4k \alpha^2 z)^{1/4\alpha^2}.
\eea
In the limit of small $\alpha$ we obtain an exponential scale factor 
\bea
a(z)=e^{-kz}.
\label{exp}
\eea
As we see from the divergent behaviour of the dilaton (which is related to a vanishing volume $V$ of the 6 extradimensions different from the orbifolded one) there is a naked singularity in the bulk and it is screened by one of the two branes (the ''hidden brane''). The remaining brane can host the standard model particles and hence will be called the ''visible brane''. Actually, as already pointed out in \cite{Zanzi:2016thx}, we have more than two branes but we will simply consider two objects: the hidden brane and a stack of N branes. In this sense we are working with two degrees of freedom in this model.

A more realistic model can be obtained by putting matter on the branes and the two parameters necessary to know the position of the branes are promoted to two moduli fields. The reader may wonder whether it is a standard process to promote two values of a coordinate to two distinct fields and how it works exactly.
In this case the Moduli Space Approximation has been used: the idea is that the evolution is not too fast, see \cite{Brax:2002nt} (remarkably some physicists don't agree with the MSA, this point is mentioned for example in \cite{Baumann:2014nda}). At this stage, forgetting about MSA, the basic idea to link a coordinate to a field is the idea of quantum fluctuations (and inhomogeneities). When we put matter on branes, we know that matter has quantum fluctuations and, therefore, the gravitational field on the brane is fluctuating too. However, the gravitational field on the brane is obtained from the gravitational field of the black hole (which depends on the distance brane-BH in the 5th dimension) and, hence, these 4D fluctuations are converted to fluctuations along the 5th dimension in harmony with the relativity of time.

The two moduli are: a dilaton $Q$ (parametrizing the position  of the center of mass of the two branes) and a radion\footnote{Do not confuse the R of radion with the one representing curvature.} $R$ (parametrizing the distance between the two branes).
$Q$ and $R$ are related to the branes' position by a couple of redefinitions. The first redefinition is
\begin{eqnarray}
\tilde \phi^2 &=& \left(1 - 4k\alpha^2 z_{visible} \right)^{2\beta}, \label{posia1}\\
\tilde \lambda^2 &=& \left(1-4k\alpha^2 z_{hidden}\right)^{2\beta} \label{posia2},
\end{eqnarray}
with
\begin{equation}
\beta = \frac{2\alpha^2 + 1}{4\alpha^2}.
\end{equation}
The second redefinition is
\begin{eqnarray}
\tilde \phi &=& Q \cosh R, \label{posib1} \\
\tilde \lambda &=& Q \sinh R \label{posib2}.
\end{eqnarray}

This 5D braneworld model can be mapped into a 5D bag exploiting a conformal transformation \cite{Zanzi:2006xr}. In particular, the naked singularity is mapped into the centre of the bag. The branes are mapped into $S^4$ spaces which are the boundaries of the bag. In other words, we obtain a spherical potential well with two $S^4$ boundaries. At this stage the local conformal symmetry is obscure, but, as we will see, this bag description of the model will be particularly useful.

Now we should mention a technicality. The singularity is a conical one when the metric $g_{\mu\nu}$ describes a compact euclidean manifold. Does this mean that we are assuming also an euclidean form of the metric $g_{\mu\nu}$ end even an eventual
compactification? It would go quite beyond a conformal transformation. The answer is that, indeed, the Euclidean picture is correct for $g_{\mu\nu}$. As far as compactification is concerned, we treat all the 4 dimensions of $g_{\mu\nu}$ democratically and a 1-point compactification should be exploited.

\section{Some results from the literature}
\label{letteratura}

In the present section we want to summarize a list of facts that will be relevant in the next sections.

\subsection{The Titius-Bode law}

In the 18th century Johann Daniel Titius von Wittenberg wrote a remarkable formula concerning the distance of planets from the Sun. Indeed, if we normalize to 10 the Earth's orbit, we can write
\be
r_n=4+3 \times 2^n
\ee
where $n$ is $-\infty$ for Mercury and 0, 1, 2,... for the remaining planets.

This initial formula has been modified in time and here we are going to discuss a formula suggested by Mary Adela Blagg in 1913, namely
\be
r_n=A (1.7275)^n [B+f(\alpha +n \beta)],
\ee
where $A$ and $B$ are positive constants while $\alpha$ and $\beta$ are angular constants. $f$ is a periodic function. More details can be found in the book \cite{Nieto:libro}.

The Titius-Bode law can be expressed, neglecting higher order corrections, in the form (see e.g. \cite{Scardigli:2005fr})
\be
r=a e^{2 \lambda_S n},
\label{TB}
\ee
where $n=1, 2, 3...$. The $\lambda_S$-parameter is related to the planetary system we are considering. In particular, for the Solar System we have
$2 \lambda_S=0.53707$ and $a=0.21363$ A.U.

\subsection{The modified Schroedinger equation}

The orbifold of time of reference \cite{Zanzi:2016thx} is a powerful tool to extract theoretical equations. An interesting example is given by a {\it modified Schroedinger equation} for quantum mechanics written in reference \cite{Zanzi:2016yrg}.\\
In the string frame (S-frame) the matter field equation is (see \cite{Fujii:2003pa})
\bea
\Box \Phi -\frac{f}{2} \phi^2 \Phi-\frac{\lambda}{6} \Phi^3 =0.
\eea
We can extract the modified Schroedinger equation following the procedure outlined in e.g. \cite{Nikolic:2012wj}\footnote{In particular we can recover the standard Schroedinger equation starting from the Klein-Gordon equation $\Box \Psi +m^2 \Psi=0$
simply writing $\Psi=\frac{e^{-imt}}{\sqrt{m}} \Psi_{NR}$, where $\Psi_{NR}$ is the non-relativistic wave function.}.
We start with a Minkowski approximation and we write (the metric signature is $-+++$)

\bea
(-\partial_t^2 + \partial_x^2 -m^2) [\frac{e^{-imt}}{\sqrt{m}} \Psi_{NR}]-\frac{\lambda}{6}  \frac{e^{-3imt}}{\sqrt{m^3}} \Psi^3_{NR}=0
\eea
where the mass is $\phi$-dependent because, by definition, $m^2(\phi)=\frac{f}{2} \phi^2$. Therefore, this equation is actually coupled to the field equation for $\phi$.
If we consider the mass roughly constant and we make the approximation of reference \cite{Nikolic:2012wj}, namely 
\bea
 \mid \partial_t^2 \Psi_{NR} \mid << m \mid \partial_t \Psi_{NR} \mid,
 \eea
then the modified Schroedinger equation is written as \cite{Zanzi:2016yrg}
\bea
i \partial_t \Psi_{NR}= -\frac{\nabla^2}{2m} \Psi_{NR} +\frac{\lambda}{12} \frac{e^{-2imt}}{m^2} \Psi_{NR}^3.
\label{mod}
\eea

This equation is manifestly non-linear. In the chameleonic model, once the dilaton is stabilized, also matter fields are fixed. This means in particular that at the level of QFT we construct a minimum in the effective potential of our theory. Near this minimum an harmonic approximation for the potential can be exploited and a linear equation can be recovered. This is the idea behind the linearization process that we will consider later in this article when we will link our equation to the equation in \cite{Scardigli:2005fr}.

\subsection{The Titius-Bode law from a quantum-like set up}

This subsection touches upon the main results of reference \cite{Scardigli:2005fr}.
We start from a Schroedinger equation of the form
\be
{\hat H}_M \psi={\cal E} \psi
\ee
where ${\hat H}_M$ is the hamiltonian per unit mass
\be
{\hat H}_M=\frac{{\bf p}^2}{2}+V(r)=-\frac{s^2}{2} \nabla^2 +V(r).
\ee
In this formula $V$ is the potential energy per unit mass, $\psi$ is the wave function, ${\cal E}$ is the energy per unit mass and $\hbar$ is replaced by a new fundamental ``constant'' $s=\hbar/m$ characterizing the planetary system we are interested in. Hence, if we consider a two-dimensional set-up where planets are localized, we can write our operators in planar polar coordinates as
\be
{\hat p}_r^2=-s^2 \frac{1}{r} \frac{\partial}{\partial r} (r \frac{\partial}{\partial r})
\ee
\be
{\hat p}_\phi^2=-s^2  \frac{\partial^2}{\partial \phi^2} 
\ee
and, therefore, the hamiltonian becomes
\be
{\hat H}_M= \frac{1}{2} [-s^2 \frac{1}{r} \frac{\partial}{\partial r} (r \frac{\partial}{\partial r})- \frac{s^2}{r^2}  \frac{\partial^2}{\partial \phi^2} ]+V(r).
\ee

To proceed further, a modification of this equation is suggested in \cite{Scardigli:2005fr} in order to establish a contact with the Titius-Bode law. In more detail, the new hamiltonian is
\be
{\hat H}_M= \frac{-s^2}{2r^2} [ r \frac{\partial}{\partial r} (r \frac{\partial}{\partial r})+ {\cal{\hat P}}_\phi^2 ]+V(r),
\label{scar}
\ee
where ${\cal{\hat P}}_\phi^2$ is, by definition, the square of the operator
\be
{\cal{\hat P}}_\phi=i e^{-i \lambda_S \partial_\phi}.
\label{Pop}
\ee

From this equation the Titius-Bode law (\ref{TB}) can be recovered. For the complete analysis the reader is referred to the original article \cite{Scardigli:2005fr}. $\lambda_S$ is present in (\ref{Pop}) and (\ref{TB}) but it is {\it not} the $\lambda_{\Phi}$ 
parameter of the MFM lagrangian. In the remaining part of the article we will simplify slightly the notation and we will write $\lambda$ instead of $\lambda_{\Phi}$.

One of our tasks in this article is to link explicitly the paper \cite{Scardigli:2005fr} with the modified Schroedinger equation (\ref{mod}).

\section{General remarks on dilatonic time}

\subsection{The Planck mass}
\label{dyngrav}

In this M-theory model, the mass scales are running as a function of the matter density of the environment \cite{Zanzi:2012du}, but a detailed discussion of the role played by quantum fluctuations is still missing: we must discuss our exponential ``magnifying lens'' in detail. Let us fill this gap. 

The Planck mass is related to the S-frame dilaton $\phi$ and to the E-frame (Einstein frame) dilaton $\sigma$ by the formula \cite{Zanzi:2012du}:
\beq
M_p= \xi^{1/2} \phi e^{-\zeta \sigma}.
\eeq
If we give an expectation value to the fields (i.e. we take into account the backreaction \cite{Zanzi:2010rs}), this implies that whenever the matter density is small, the Planck mass (and also the mass of matter particles) becomes very small. However, this is only the background value of the Planck mass. Indeed, the dilaton is characterized not only by its expectation value but also by its quantum fluctuations.  What is the size of these fluctuations? The scale factor in the E-frame is
\beq
a_E = a_S e^{\zeta \sigma},
\eeq 
where $a_S$ is a constant related to the stabilization of the dilaton in strong coupling (see the next paragraph).
Consequently, this exponential scale factor plays the role of a ``magnifying lens'' for the quantum fluctuations. Hence we are led to a scenario where not only the background value of the dilaton but also its quantum fluctuations are a decreasing function of the matter density.  

It is time to analyze the gravitational sector of the lagrangian in more detail. What is the value of the Planck mass in this model? We write the Planck mass as
\beq
M_p \propto [<\phi>+\phi_f]e^{-\zeta \sigma}
\eeq
where we explicitly separated the expectation value of the field by its quantum fluctuation and, on general grounds, we have $\sigma=\sigma_b(t)+\sigma_f(x)$ (i.e. $\sigma_b(t)=<\sigma>$). If we shift the value of $<\sigma>$ from small to large values (i.e. we run from the UV to the IR region), following the cosmological expansion, then the background value of the Planck ''mass'' and its quantum fluctuations are subjected to {\it opposite} rescalings. Indeed, the background is exponentially suppressed (and this drives the cosmological expansion or, equivalently, the renormalization of the bag radius \cite{Zanzi:2006xr}), while the quantum fluctuations are exponentially enhanced by our ``magnifying lens''. 
The final result in the IR (i.e. large $<\sigma>$) is that 
\beq
M_p= \phi_f \mid _{<\sigma>=0} e^{\zeta \sigma} e^{-\zeta \sigma}= \phi_f \mid _{<\sigma>=0}.
\label{growth}
\eeq
Some comments are in order regarding this formula. The exponential growth $e^{\zeta \sigma}$ of $\phi_f$ in \ref{growth} is due to the exponential behaviour of the scale factor: during the cosmological expansion, length scales and fluctuations are stretched to larger and larger values by the exponential magnifying lens.

 In the UV\footnote{UV means short length scales. This situation corresponds to the strong coupling of the theory.} the expectation value (i.e. the background value) of the Planck ``mass'' is the dominant one (because there is no exponential magnifying lens), scale invariance is abundantly broken and the Einstein-Hilbert term is recovered. On the contrary in the IR the quantum fluctuations are dominant in the Planck ``mass'' (because the ``magnifying lens'' is operative), scale invariance is restored and the Einstein-Hilbert term is replaced by a non minimal coupling term of the form
\beq
{\phi}_f^2 R.
\eeq

Indeed, $<\phi>$ is not subjected to the exponential ``magnifying lens'' and, therefore, in the IR it can be safely neglected with respect to the quantum fluctuations $\phi_f$. Remarkably this construction is in harmony with the orbifold of time telling us that we can use a non-minimal coupling term in the IR. 
Let us further elaborate these points. Let us think to the dilatonic time dimension {\it globally}. This means that we consider not only the post big bang phase, but also the pre big bang one. The VEV of $\phi$ has been fixed by the gauge fixing procedure and, therefore, it cannot change anymore. When we use $\sigma$ as our dilaton field in the post big bang phase, we are taking into account the fact that a gauge-fixing for $\phi$ has been done.

There are a number of consequences of this magnifying lens. In particular, the Planck ``mass'' is entirely dominated by quantum fluctuations in the IR. The same result is true also for matter fields, because $\Phi \simeq \phi$. The final outcome is a {\it quantum universe in the IR region}. However, the classical world we are used to is recovered on the one hand locally (e.g. in Zurich), where the matter density is large, because the Planck mass is dominated by its classical background value (i.e. we switch off the exponential magnifying lens for the quantum fluctuations and hence we are left with the background value of the fields), on the other hand on planetary distances, because we assume that quantum fluctuations are stretched by the exponential lens on length scales much larger than 10 astronomical units and, therefore, physics classicalizes in the solar system today. In other words, a fluctuation with wavelength much larger than the scale where we perform our measurements is unobservable. We can summarize this situation with a $\sigma$-dependent de Broglie wavelength.

 The next question is whether this gravitational theory is in harmony with planetary motion or not. As already mentioned in \cite{Zanzi:2016thx}, we can write the string mass as
\beq
M_S^2 \propto \rho k_{11}^{-2/3}
\eeq
while the Newton ''constant'' is given by \cite{Zanzi:2016thx}
\beq
G_N=\frac{k_{11}^2}{16 \pi^2 V \rho}.
\eeq 
Therefore the gravitational coupling $\alpha_{G} \simeq G_N M_S^2$ (we are using the string mass as a UV cut off) is switched off in the IR region (where $V$ is large). Now let us  analyze gravity in the solar system. When we consider the interaction Earth-Sun (or Jupiter-Earth...), it is true that we evaluate the gravitational field in the position of the Earth but this is not enough to exclude the presence of large distances in the theory: the gravitational field felt by the Earth depends on the distance Earth-Sun and this introduces into the theory a much larger length scale. This implies a suppression of the background value of the Planck ``mass''. Is this a problem from the phenomenological point of view? Happily, even if the background Planck ``mass'' is small, gravity remains dynamical in the IR because quantum fluctuations are important. Now to the point: does a small $\alpha_G$ in the IR clash with the constraints on the orbital motion of planets? It is common knowledge that gravity is the weakest of all interactions. Hence, in the solar system, gravity is very weak but its effect on the motion of planets is relevant simply because on a planet we have a coherent superposition of small effects. On the other hand, in the solar system the dilaton is very light and the standard argument to keep under control phenomenological constraints on fifth-force effects in the solar system is to exploit a thin-shell mechanism (for a discussion of the thin-shell mechanism, see e.g. \cite{Zanzi:2015cch}). The final result in the solar system is a scalar-tensor theory of gravitation with a very light chameleonic dilaton whose effects on planets are screened by the thin-shell mechanism. This is one of the good aspects of taking into account quantum fluctuations: gravity is still dynamical in the IR and phenomenological constraints can be faced more easily.

Let us further discuss the conformal anomaly. If we consider only the expectation values of the fields, then we obtain the matter density starting from the anomaly-induced term $M_p {\tilde \Phi_*}^2 \sigma$ (when we write ''anomaly-induced term'', we mean that there is an interaction term which is induced by the anomaly, see e.g. \cite{Fujii:2003pa, Zanzi:2010rs}). What is the role played by quantum fluctuations? We know that these fluctuations are enhanced in the IR region and, therefore, in the IR we have an interaction term between dilaton and matter. This term does not contain expectation values of fields and, hence, it does not clash with the restoration of conformal invariance in the IR. The situation is totally analogous to the Yukawa terms in the SM. Indeed, we can start considering an interaction term of the form Higgs-fermion-fermion and if we give an expectation value to the Higgs field, we are left with a mass term for the fermion (we don't have an interaction term anymore). In the MFM, we build the matter density taking the expectation values of our anomaly-induced term, but, on the contrary, quantum fluctuations induce only an interaction term for the dilaton in the IR. This interaction term might be the source of potentially detectable deviations from standard cosmology.

One last comment is in order. One might say that we cannot switch off the interactions when the matter density is small, because at 300 Mpc the matter density is small but gravity is not absent. Indeed, we know that at 300 Mpc gravity is still present even if the matter density is basically critical. This fact we mentioned last can be explained in our model. Indeed, our exponential magnifying lens for quantum fluctuations produces a non-minimal coupling term for gravity in the IR and, hence, infrared gravity is still dynamical. Even if the parameter $\alpha_G$ is very small, gravitational interaction is the result of a coherent superposition of a large number of small contributions and, consequently, gravity is detectable also at 300 Mpc.

\subsection{Dilaton stabilization in strong coupling}

Let us start considering heterotic string theory in the framework of Pre-Big-Bang (PBB) scenario (see for example \cite{Gasperini:libro} and related references). Our intention is to work with the 5D bag model mentioned above. The theory has only a global scale invariance (in the IR), not a local one, therefore, the bag description might be considered only as a useful tool to perform calculations. We will come back to this issue later. 

Our first purpose is to stabilize the dilaton in strong coupling with a non-approximated potential. This stabilization is crucial to keep under control the cosmological constant (this point is clarified in \cite{Zanzi:2010rs}).  We interpret this bag model as a black hole with its horizon. 
Since a 4D black hole has $S^2$ as horizon, a 5D black hole has a 3-dimensional horizon. This is a first clue of the presence of a second timelike dimension in our model. It is common knowledge that inside the horizon of a black hole, the radial dimension becomes timelike (this is true already for a Schwarzschild black hole). Notice however that the orbifold of time is not necessarily linked to the black hole interpretation. In string theory we can exchange space with time also without a black hole. It is necessary that spacetime is euclidean though. Therefore, this scenario is taking shape: the radial coordinate of the bag is an orbifold of time, while on the boundary of the bag we have a second timelike dimension which gives us a closed timelike curve. Let us postpone the discussion of 2-time physics and let us start considering only the orbifold of time (see \cite{Zanzi:2016thx}). We will call ''dilatonic'' the orbifolded time (see paragraph \ref{5c}). Before we impose the orbifold parity we have only a circle of time. The upper half circle corresponds to the PBB phase (negative time), the lower half circle corresponds to the post-big-bang phase (positive time). With this peculiar orbifold of time our branes become S-branes (see \cite{Gutperle:2002ai}) and their dynamical evolution, parametrized by the moduli fields, takes place along the timelike coordinate. For this reason the dynamical evolution of the moduli is related to the flow of time. If we change the value of a modulus parametrizing the position of a S-brane, then we change the value of time on that particular brane. This information comes directly from the geometry of the problem and, for this reason, we will call this statement our ''{\it geometric argument}''. Hence, in order to stabilize the dilaton, we must answer this question: how does time evolve? We trade the problem of dilaton stabilization with the problem of a quantitative control on the flow of time. 

The simplest possibility is to imagine a point on the circle of (dilatonic) time moving with constant angular velocity. The relativity of time tells us that we must split our analysis considering separately an observer near the black hole and an observer far away from the black hole. This point must be further elaborated. In the strong coupling regime of heterotic theory an 11th dimension is present. The size of this 11th dimension is much smaller than a cosmological scale but much larger than the Planck length. It is an intermediate length scale. In our 5D model, the 5th dimension corresponds to the original 11th dimension. We infer that the size of the 5th dimension is intermediate in strong coupling. In other words, an observer near the black hole perceives an intermediate 5th dimension and, hence, the fluctuations of the dilaton are constrained to remain inside this scale. However, the size of a dimension is observer-dependent and these considerations do not forbid the existence of an asymptotic observer located at a cosmological distance from the black hole. The trick is to apply the relativity of time (analogously, a relativistic observer would be able to travel along our observable universe in only 45 years, not 13 Gy - see the majestic book \cite{Gourgoulhon:2013qva}, p. 53). Consequently, the stabilization of the dilaton in strong coupling is automatic for an asymptotic observer who wants to keep under control the cosmological constant. From the standpoint of an asymptotic observer, strong coupling scales are microscopic and time near the black hole is frozen. Hence, {\it for an asymptotic observer, the dilaton is stabilized in strong coupling and this result is non-approximated}. 

\subsection{Background dilaton as a cosmological time}
\label{5c}

Since the dilaton is stabilized in strong coupling, the string frame scale factor $a_S$ is constant because the bag radius is constant and we live on a portion of a $S^4$ brane (this statement is non-trivial because the presence of a closed timelike curve is forbidden in $S^n$ spaces with $n>1$ and, moreover, we will discuss a fibre bundle structure for spacetime; at this stage we simply accept this bag scenario, because these issues will be clarified later). After the conformal transformation to the E-frame, the new scale factor will be $a_E=a_S e^{\zeta \sigma}$. Exploiting our geometric argument of the previous paragraph, we know that in the pre big bang the dilaton was running towards the strong coupling, because time was oscillating towards the black hole (i.e. the point representative of time was moving along the upper half circle in a counterclockwise sense). Hence, the orbifold of time guarantees the running of the dilaton to the weak coupling in the post big bang. We now identify the exponential scale factor of the MFM $a_E=a_S e^{\zeta \sigma}$ with the exponential scale factor of heterotic-M-theory valid for small values of $\alpha$, namely $a(z)=e^{-kz}$. With this identification, the region near the singularity corresponds to the strong coupling of the theory (as required) and $\sigma \propto z$. Remarkably, when we exchange the fifth coordinate with time, this formula we mentioned last seems to suggest that
\beq
\sigma_b(t) \propto t.
\eeq
 This is in harmony with our geometric argument linking moduli and time. Therefore, we suggest to identify the background dilaton $\sigma_b$  with our cosmological time coordinate. The dilaton oscillates in time but it can also be identified with time. Hence, our geometric argument tells us that the dilaton is a quantum of time, namely a chronon (see also \cite{Caldirola:1977fu} and related references).

A dilatonic time has a number of non-trivial consequences. First of all, the stabilization of the dilaton near the big-bang (i.e. the stabilization of the bag) can be interpreted as a Dirichlet condition imposed on the cosmological time. When the dilaton is stabilized we fix the cosmological time. After the stabilization has taken place, the post-big-bang time evolution can be described by a renormalization of the Planck mass to smaller values: the value of the dilaton is fixed in planckian units, but the Planck mass is decreasing and, hence, the bag is expanding. This picture is taking shape: the post-big-bang expansion can be interpreted as a renormalization of the Planck mass to smaller values but, during the expansion, the cosmological time remains fixed. In this way, as already mentioned above, our 4-branes become the S-branes of \cite{Gutperle:2002ai}, because a Dirichlet condition is imposed on the cosmological time.

This idea of cosmological time evolution as renormalization can be analyzed also at the level of the effective action in 4D. Indeed, the renormalization of masses and gauge couplings in the 4D theory suggested by the MFM is related to a shift of the value of $\sigma$ (which is running to large values) and, consequently, the renormalization plays the role of a time evolution also at the level of the effective action. However, we can infer one more piece of information from this effective action analysis. Indeed, the Chameleonic Equivalence Principle (CEP, see \cite{Zanzi:2015evj}) is telling us that a shift in the value of $\sigma$ is related to a variation of the gravitational field and, therefore, we infer that the renormalization of the parameters of the 4D theory is due to gravity. In other words, {\it the time evolution of the parameters (gauge couplings and masses) of a 4D gauge theory is determined by a variation of the gravitational field.} The renormalization is ''exchanged'' with the time evolution and this is fully compatible with the exchange of the 5th coordinate with time (remarkably, in AdS/CFT it is a standard procedure to interpret a shift in the 5th dimension as an RG-running in 4D).

The careful reader might think that a Dirichlet condition on time is not so intuitive. However, if we imagine to live on the horizon of the 5D black hole during our entire cosmological history then we find a very simple explanation: on the horizon we have an infinite time dilation and our time is frozen from the standpoint of an observer at infinity. In this way, during the cosmological expansion, the renormalization of the Planck mass leads to a dynamical behaviour of the horizon and, hence, to a dynamical behaviour of our brane.

One last remark is in order. When we write the scale factor as $a \simeq e^{\zeta \sigma}$ and we identify the background dilaton with time, then we are led to a de Sitter universe which naturally clashes with cosmological observations. However, this is not a ''standard de Sitter'' evolution because we have to take into account the presence of the black hole and of string dualities. Indeed, on the one hand, the black hole will modify the evolution of our dilaton-chronon, on the other hand, S-duality will identify the weak coupling with the strong coupling and for this reason, as we will see, the Dark Energy phase will be dual to the inflationary phase. These issues will be discussed later in more detail.\\

{\bf Discussion:}\\

\begin{itemize}
\item {\it Question: Here we are identifying a field with a coordinate. Does this mean that we are replacing the field with its expectation value? What is the meaning of such identification? Is it a standard procedure?} 

{\it Answer:} We write $a(z)$ in two different ways (following the MFM and following heterotic-M-theory) and these two ways must coincide because the lagrangian of the MFM can be obtained from heterotic-M-theory (see \cite{Zanzi:2016thx}). This is the main argument, but we can add some comments. When we analyze the wave function of bulk matter, we find that the exponential wave function is localized towards the hidden brane (see \cite{Zanzi:2016thx}). We can use this wave function to parametrize the position $z$ in the bulk. The vev of the matter field (squared) is related to the number density of particles, hence, to the wave function and, therefore, to $z$. After all, there are some clues of this identification already in standard QFT. Indeed, when we have classical external fields coupled to currents, we obtain additional Feynman vertices carrying position labels $x, y,...$ over which we do not integrate (see \cite{Weinberg:libro} p. 287). The integration over $d^4 x$ is related to quantum fluctuations and since the external fields are classical, there is no integration. The analogy with the path integral is clear and this seems to suggest a formal link between $d^4 x$ and ${\cal D} \phi$. One last comment is in order. Later in this paper we will analyze the time operator in connection to the extended Mukhanov-Sasaki equation and this seems to explain, at least partially, the meaning of the identification between a field operator and a coordinate. 
\item {\it Question: It would not be surprising to identify $\sigma$ with time if $\sigma$ was a classical field depending on $t$ only. This is what happens with the background dilaton $\sigma_b(t)$. However, in general, time must be also a function of space coordinates because the gravitational field is not homogeneous. Moreover, is it possible to show that $\sigma$ is a monotonic function of time, at least for a given solution? This is what happens with $\sigma_b(t)$ in the deep IR, but when quantum fluctuations become non negligible, why can we keep the identification? Of course this identification would give a quantum nature to the time, however it would also mean that the identification is not merely occasional but is a consequence of a deeper meaning of $\sigma$.}

{\it Answer:} If you identify $\sigma_b$ with time and this field is classical, why don't we move to the quantum level? Typically, when we quantize a field we start from a classical field theory and then we quantize it. In this case, we start with the identification of $\sigma_b$ with the cosmological time and then we quantize it. 
\end{itemize}

\subsection{Coherent states}
\label{coherent}

As already mentioned above, the cosmological evolution is determined by the motion of the dilaton-chronon. Indeed, the dilaton parametrizes the position of the brane. Here is where the notion of the aforementioned identification dilaton-coordinate comes in. $\sigma$ does not really measure the radial coordinate but the radius of the bag and its change  specifies in which instant the dirichlet condition is imposed on time. However, if the expectation value of the dilaton tells us the instant of time on the brane, we call the dilaton ''time''. Equivalently, if the expectation value of the dilaton tells us the position of the brane, we call ''dilatonic'' the radial direction.

 If we consider the simplest possibility, namely a constant angular velocity on the circle of (dilatonic) time, then the $\mathbb{Z}_2$ parity produces an harmonic oscillation of time and, exploiting our geometric argument, we expect the dilaton to oscillate from the weak coupling (well before the big bang) to the strong coupling (near the big bang) and then back to the weak coupling (in our low redshift universe). Consequently, we will exploit coherent states to describe the oscillatory motion of the chronon in the bulk. During this oscillation, when the dilaton reaches the non-perturbative region, we have 11 dimensions and, at intermediate energies, we have the 5D model with two branes described in section \ref{model} in the framework of heterotic-M-theory. As already mentioned above, we have two moduli in the effective action of the SUGRA model, namely the dilaton and the radion (as we will see, they are both timelike coordinates). However, the lagrangian of the MFM is telling us to consider only one modulus. How can we obtain this result? As far as the radion is concerned, we will simply assume that the potential of the theory will give us a chameleonic radion with a large (small) value at small (large) redshift. This behaviour of the radion is in harmony with the analysis presented in \cite{Brax:2004ym} (as we will see in section \ref{sit}, the chameleonic behaviour of the dilaton and of the radion is an effective description of the relativity of time in a gravitational field). When we work with two timelike chameleons, we are free to redefine the time in a convenient way: we can define ''the dilaton'' as the timelike coordinate in the 2-dimensional manifold of time linked to our everyday time. The ''orthogonal'' timelike coordinate is naturally the radion and it is stabilized by some UV dynamics simply by assumption. A word of caution is necessary regarding this stabilization of the radion. Even if in our effective lagrangian the radion is stabilized (and hence integrated out), this does not mean that there are no fluctuations for the radion in our everyday life. These fluctuations are extremely small but they are present and, moreover, they are sensitive to the matter density of the environment. Therefore, {\it if we change the matter density of the environment, we modify the value of the dilaton and also of the radion}. For a fixed $\Delta \rho_m$, the dilatonic variation is much larger than the radionic one. However, this small radionic shift will be crucial later. At this stage, we keep on working with a single (dilatonic) time dimension and we postpone to the following sections a more careful investigation of 2-time physics in our model.

The harmonic oscillation of (dilatonic) time is the simplest possibility but it does not take into account the effect of the gravitational field of the 5D black hole. We can study the time dilation in a gravitational field using energy conservation. In our 5D model the situation is similar: the presence of the gravitational field of the black hole produces a relativity of time which modifies the flow of time. In this way we have an unperturbed time which flows in a harmonic way on the orbifold and corresponds to a constant angular velocity on the circle. However, we can take into account the effect of the black hole on time exploiting the analogous calculation done with the charged HO perturbed with a time-dependent electric field.

The idea is to build a dictionary between exercise number 6 p.636 of \cite{Cohen:2005cra} (whose notations we follow here) and the gravitational problem of our theory we are interested in. The dictionary is presented in table \ref{tab1}.

\begin{table}[t]
\center
\begin{tabular}{| c|c | }
\hline
               {\bf Problem with t-dependent electric field} & {\bf Problem with z-dependent gravitational field}  \\
\hline

Time is a classical parameter      & $z$ is a classical parameter   \\
\hline
$X$ is a quantum operator       & ${\hat T}$ is a quantum operator      \\
\hline
Electric field ${\cal E}(t)$        & Gravitational field of the BH ${\cal G}(z) $     \\
\hline
Electric charge $q$        & Unrenormalized Planck mass $M_p^{unren}$     \\
\hline
   Gaussian wave packet gives the unperturbed HO  & Gaussian wave packet describes time without BH   \\
\hline
Charged particle      & Massive dilaton   \\
\hline

\end{tabular}\\[1cm]
\caption{Dictionary between the exercise of \cite{Cohen:2005cra} and our oscillating dilaton.}
\label{tab1}
\end{table}

Some remarks are necessary: 
\begin{itemize}
\item the exchange of the fifth coordinate with (the dilatonic) time is formally obtained starting from the exercise of \cite{Cohen:2005cra} and imposing a shift of the origin of the 5th coordinate in the MFM. Indeed, the origin of time $t=0$ (i.e. the big bang) coincides with one of the fixed points of the orbifold. Analogously, the position of the black hole in the 5th dimension will be defined to be $z=0$ in this paragraph and {\it not} $z=\frac{1}{4k \alpha^2}$ like in the standard literature.
\item The $\alpha$ parameter evaluated for the charged HO with electric field will give us (in the dual problem) the behaviour of time in the presence of the black hole (i.e. taking into account the gravitational time dilation near the black hole). 
\item The $\omega$ parameter of the oscillator is replaced by $\omega_u$, the frequency characterizing the unperturbed motion of time with constant angular velocity. 
\item The charged particle of the exercise is replaced with a massive dilaton oscillating along the orbifold of time.
\item The instant of time $T$ where we switch off the electric field is replaced by a distance $z^*$ from the black hole. For $z>z^*$ the gravitational field of the black hole can be neglected, but this is never the case for an observer near the black hole.
\end{itemize}

To proceed further, we are going to split the orbifold of time in four sectors: 1) from the center of the orbifold to the black hole in the PBB phase; 2) from the black hole to the center of the orbifold in the post big bang phase; 3) from the center of the orbifold to the second orbifold fixed point in the post big bang phase; 4) from the second orbifold fixed point to the center of the orbifold in the PBB phase.

Sector one will be analyzed exploiting coherent states. Once the dynamical evolution of the Universe is known in sector one, we can use the orbifold of time to infer the evolution in sector two. Then we can use S-duality to map the strong coupling of the post big bang (sector two) in the weak coupling of the post big bang (sector three). Finally, sector three can be mapped into sector four exploiting the orbifold of time. In this way, the acceleration of the chronon towards the black hole in sector one provides an inflationary phase in sector two and the Dark Energy phase in sector three. Consequently, once the cosmological evolution is known in sector one, we will infer the complete cosmological evolution. We will say more about this point later.

In this way we can exploit the results of the exercise to write our new Hamiltonian for sector one as 
\be
H= \frac{p_{T}^2}{2M_p^{unren}} + \frac{1}{2}M_p^{unren} \omega_u^2 {\hat T}^2 -M_p^{unren} {\cal G}(z) {\hat T}
\label{Ham}
\ee
where the gravitational field of the black hole can be written in this extradimensional set-up as 
\be
{\cal G}(z)= G_N M_{BH}/z^4
\ee
exploiting the Gauss' law.
Naturally we write
\be
<{\hat T}>(z)=\sqrt{2 \hbar/(M_p^{unren} \omega_u)} \{ \alpha_R cos[\omega_u (z-z_*)]+\alpha_I sin[\omega_u (z-z_*)]\}
\ee
and
\be
\alpha(z)=e^{-i \omega_u z}[\alpha(z=0)+i \int_0^{z} ds \lambda(s) e^{i \omega_u s}],
\label{formuletta2}
\ee
where
\be
\lambda(z)=\sqrt{\frac{M_p^{unren}}{2\hbar \omega_u}}{\cal G}(z).
\ee
A numerical analysis of the parameter $\alpha$ is definitely necessary. \\

{\bf Discussion:}\\

{\it Question: What is the physical meaning of this Hamiltonian? It is obtained by the analogy with the oscillator in the electric field, 
but now the electric charged particle is replaced by the oscillating time. Why 
the Hamiltonian should be written in this way? Is it an evolution with respect to the second time? Why the effect of the black hole should be so simple? A more complicated contribution should be present unless we are very far from the black hole.}

 {\it Answer:} It is the hamiltonian of the dilaton but, as we will discuss later, the oscillation of the particle is only an effective way of summarizing S-duality and orbifold parity. Therefore the structure of the harmonic oscillator is a direct consequence of S-duality and orbifold parity. At this stage it seems that this is an evolution with respect to the second (radionic) time. The radion is stabilized but, as we will further discuss later, its quantum fluctuations are not absent and they provide a fundamental timelike ''tick''. Hence the kinetic term of the dilatonic chronon in the hamiltonian describes the velocity of the dilatonic time with respect to the radionic time. As far as the black hole is concerned: IR is dual to UV exploiting S-duality and, hence, if we accept this Hamiltonian at very large distances from the black hole, then we should accept it also at very small distances. In principle the intermediate region remains as an open problem.

\subsection{A first look at $T^2$ geometry}

In a future line of development we will analyze the $\alpha$ parameter of formula \ref{formuletta2} for sector 1 with the help of a computer. However, one problem must be faced: the integral  in \ref{formuletta2} diverges in $z=0$ and, hence, we need a UV cut off for our integral to be well-defined. What kind of cut-off should we use? In other words, what is the shortest distance from the singularity that the theory can allow? To answer this question, we start pointing out that the theory we are considering is effectively 2-dimensional. We are working with the 5th coordinate $z$ and with the time $t$. Actually this two-dimensional construction is even more useful. Indeed, as already mentioned above, in our model there are two timelike directions and we will build a torus of time. However, at this stage, we keep on working with one timelike and one spacelike dimension in order to add gradually new elements into the model. 2-time physics is described later.

In order to discuss the problem of the UV cut off for the integral, let us exploit a $T^2$ geometry. Now we build this torus. Our two dimensions are the 5th coordinate $z$ compactified on a $S^1$ circle and the time $t$ with the geometry of a $S^1/\mathbb{Z}_2$ orbifold. Consequently, the geometry of our 2D theory is an annulus where the internal $S^1$ circle corresponds to a spacelike dimension where the theory is strongly coupled, the external $S^1$ circle corresponds to a spacelike dimension where the theory is weakly coupled and the radial direction corresponds to time (i.e. the orbifold of time). We are led to this scenario: the annulus can be interpreted as a 1-loop diagram for the ''open string'' that we call time (as we will see later, from the standpoint of chronon physics there is no difference between spacelike and timelike, therefore in a M-theory language this string can be interpreted as a wrapped M2-brane). If we {\it assume} conformal invariance in this worldsheet theory (the analysis of the critical/non-critical nature of this string theory is left for a future work), namely the 2D conformal invariance of the worldsheet of the ''string'' that we call time, we can map the annulus into a cylinder. One basis of the cylinder is a closed spacelike curve where the theory is strongly coupled, the remaining basis of the cylinder is a closed spacelike curve where the theory is weakly coupled. The dimension parallel to the axis of the cylinder is timelike. If we exploit S-duality, we can exchange the weak coupling with the strong coupling and, consequently, we can map the cylinder into a $T^2$-torus. To be more explicit, a modular S-transformation would correctly summarize the operation on the torus (see e.g. p. 117 of \cite{Blumenhagen:2009zz}). In other words, S-duality allows the cylinder to be closed into a torus. It is common knowledge that the moduli space of the $T^2$-torus is non-trivial. In particular the complex structure of $T^2$ is defined by a pair of complex numbers, $\omega_1$ and $\omega_2$, modulo a constant factor and $PSL(2, Z)$. If we call $H$ the set of complex numbers with positive imaginary part, then in order to remove the constant factor we can introduce the modular parameter $\tau\equiv \omega_2/\omega_1 \in H$ and we can specify the complex structure of $T^2$. We can take 1 and $\tau$ as generators of the lattice. Needless to say, not all $\tau$ are independent modular parameters and we are led to the quotient space $H/PSL(2,Z)$. It is precisely in this way that we obtain our UV cut-off. Indeed, an inspection of the quotient space tells us that the absolute value of $\tau$ must be greater than 1 and, hence, the radius of the inner $S^1$ space of the annulus cannot be less than $1/(2\pi)$ (in unrenormalized planckian units). Now that we found our UV cut-off the integral of formula \ref{formuletta2} is well defined and we can proceed with the numerical analysis for sector one. The remaining sectors are linked to sector one exploiting the orbifold of time and S-duality. 

As already mentioned above, we postpone the numerical analysis to a future publication.

\subsection{A space-independent hamiltonian and its uplifting}
\label{sit}

As we know the dilaton is a chameleon in this model \cite{Zanzi:2010rs} and this peculiar behaviour of the mass must be recovered from the curvature of the potential. To better illustrate this point we can define an effective angular frequency $\Omega$ combining the unperturbed oscillator with the gravitational perturbation. However, we exchange the 5th coordinate of the gravitational potential with time and, hence, we obtain a {\it space-independent} hamiltonian. Therefore, the effective angular frequency becomes
\be
\Omega^2 (T)=\omega_u^2 + \frac{2G_N M_{BH}}{T^5}
\ee
where $T$ is the eigenvalue of the (dilatonic) time operator and $\omega_u$ is the angular frequency of the uniform circular motion of the dilaton-chronon in the absence of gravity. 
This $\Omega$-parameter controls the curvature of the potential and, hence, the effective mass of the dilaton.
Remarkably, we flipped the sign before the second term: this sign flip is allowed by the orbifold of time (because we identify $T \longleftrightarrow  -T$) and in this way we recover the chameleonic behaviour of the dilaton. Indeed, far away from the black hole (i.e. in the large-$T$ region) the mass is small because the curvature of the potential is small and the vacuum energy is small, but, on the contrary, near the black hole the gravitational field produces a diagonal displacement of the parabola approximating the stabilizing potential near its minimum and, hence, the mass of the dilaton is larger and the ground state energy is not small anymore. Needless to say, this chameleonic mass is in harmony with the work done by the gravitational field. Therefore, interestingly, {\it the chameleonic behaviour of the dilaton is an effective description of the relativity of time in a gravitational field.} In other words, {\it the renormalization of the Planck mass is interpreted as the effect of the perturbation of the black hole}. In this theory, the gravitational field of the brane (i.e. the gravity we feel in our everyday life) is directly obtained from the gravitational field of the 5D black hole: brane gravity is a ''slice'' of the bulk gravity and, in this sense, the model is strongly holographic. However, as we will see later, the orbifold of time will produce some peculiar phenomena on the brane which are reminiscent of Newtonian gravity.

This sign flip must be further analyzed. Let us start from the hamiltonian \ref{Ham} and let us exchange the 5th coordinate with time. In this way we can write two space-independent hamiltonians:
\be
H_{\pm}= \frac{p_{T}^2}{2M_p^{unren}} + \frac{1}{2}M_p^{unren} \omega_u^2 {\hat T}^2 \pm M_p^{unren} G_N \frac{M_{BH}}{T^4} {\hat T}.
\label{hpiumeno}
\ee
where $T$ is the eigenvalue of the (dilatonic) time operator ${\hat T}$. The $\mathbb{Z}_2$ parity of the orbifold tells us that we can identify $H_+$ with $H_-$. This point is non-trivial because the term
\be
M_p^{unren} G_N \frac{M_{BH}}{T^4} {\hat T}
\ee
seems to break the $\mathbb{Z}_2$ symmetry explicitly. However, an explicit breakdown of the $\mathbb{Z}_2$ symmetry is forbidden by the geometry of time. The situation is reminiscent of spontaneous symmetry breaking.

 Let us consider the simplest possibility (without sign flip), namely
\be
H_-=\frac{p_{T}^2}{2M_p^{unren}} + \frac{1}{2}M_p^{unren} \omega_u^2 {\hat T}^2 - M_p^{unren} G_N \frac{M_{BH}}{T^4} {\hat T}.
\label{H}
\ee
and let us imagine to work with the post-big-bang phase where time is positive ($T>0$). The kinetic term $\frac{p_{T}^2}{2M_p^{unren}}$ and the harmonic term $\frac{1}{2}M_p^{unren} \omega_u^2 {\hat T}^2$ give a positive contribution to the ground state energy. However, the last term, namely $- M_p^{unren} G_N \frac{M_{BH}}{T^4} {\hat T}$ gives a negative contribution to the ground state energy and, moreover, the more we are close to the big bang (i.e. small-$T$ region), the more this contribution becomes important. This is a bad news because a dS ground state is necessary to obtain a chameleonic dilaton in the MFM (see \cite{Zanzi:2010rs}). We have an uplifting problem. Naturally, this is not the only model where an AdS ground state must be uplifted. For example, in the KKLT model \cite{Kachru:2003aw}  an AdS ground state is uplifted exploiting anti-D3-branes and many other models have been discussed in the literature. Remarkably, in our theory, the uplifting of the ground state is obtained exploiting the orbifold of time: we have {\it an orbifold-induced uplifting}. Indeed, if we flip the sign in the last term of \ref{H}, the contribution to the ground state energy becomes positive for $T>0$. Hence, we are led to the following scenario:  the stabilizing potential for the dilaton is approximated near the minimum with a harmonic potential (a parabola); there is a diagonal displacement of this parabola because the gravitational field of the black hole perturbs the system; when the time is positive (i.e. post-big-bang) the correct hamiltonian to obtain a dS ground state is $H_+$; when the time is negative (i.e. pre-big-bang) the correct hamiltonian is $H_-$; the sign flip $H_-\longleftrightarrow H_+$ is allowed by the orbifold of time and, in this sense, the uplifting is due to the orbifold of time. To the best of our knowledge, this orbifold-induced uplifting has never been discussed before in the literature. Naturally, the ground state energy is not invariant under the sign flip. In this scenario, the vacuum energy is always positive and it is a decreasing function of the distance from the black hole (in harmony with the chameleonic description of the dilaton of reference \cite{Zanzi:2010rs}). As we will see, the deep UV region (i.e. transplanckian energies) will require additional comments.

One more comment is in order. We might wonder whether the sign flip $H_- \longleftrightarrow H_+$ means that we can exchange attractive gravity with repulsive gravity. The best interpretation, at this stage, seems to be that a black hole in the PBB phase becomes a white hole in the post big bang phase.

\subsection{The meaning of the time operator}
\label{timeop}
We introduced a time operator, but its physical meaning is still obscure. After all in quantum mechanics time is not an operator. In this section we will show that the time operator is the comoving curvature perturbation.

We will start considering a short time interval just after the bounce (i.e. all the branes are very close to the black hole). The string dilaton and the string frame matter field have been stabilized and we can integrate them out when we construct our effective action. Indeed, our intention is to exploit the effective theory of inflation of reference \cite{Cheung:2007st} (see also appendix B of reference \cite{Baumann:2014nda}) where a field $\pi$ is introduced in the theory as the Goldstone boson of broken time translation. Remarkably, our field $\sigma$ is the Goldstone boson of broken scale invariance, but the exchange of the fifth coordinate with time allows the identification of $\sigma$ with the $\pi$ field: $\sigma$ is now the Goldstone boson of broken time translation. 

In order to exploit the scenario of \cite{Cheung:2007st} we must avoid a pure de Sitter universe. In the MFM we have an E-frame scale factor given by
\beq
a_E=a_S e^{\zeta \sigma}
\eeq
where $a_S$ is related to the stabilization of the S-frame dilaton and $\zeta$ is a constant of order one. This is a de Sitter universe with $H=\zeta$. As we will now prove, $\zeta$ is actually a field encoding the quantum fluctuations of the S-frame dilaton. Let us clarify this issue. The S-frame dilaton $\phi$ is stabilized but its UV quantum fluctuations can be integrated out only below the UV cut-off scale characterizing the effective action (where $\sigma$ is exploited to take into account the gauge fixing). When we consider a universe just after the bounce, we are considering energies just below the cut-off (because the cut-off is simply the scale where we break spontaneously time translations). Consequently, the effective theory is pushed near the UV cut-off and the process of integrating-out is not completely efficient: the quantum fluctuations of the S-frame dilaton are perceived at the level of the effective action and they can be summarized by a varying $\zeta$ field. These variations are characterized by a very large mass and hence by a very small wavelength (i.e. a very large frequency scale). We infer that the time variation of this field is very fast and, hence, ${\dot H}$ is not small. This comment has a number of consequences:
\begin{itemize}
\item We do not have slow roll inflation (the parameter $\epsilon$ of \cite{Baumann:2014nda} is not small).
\item We can understand better the origin of the mass of $\sigma$. In reference \cite{Zanzi:2015evj} an equivalence between gravitation and conformal anomaly has been studied as a consequence of the chameleon mechanism. Now we see, following equations (B.54-B.55-B.56) of \cite{Baumann:2014nda}, that the origin of the mass of the dilaton is the mixing of $\sigma$ with gravitational fluctuations. This mixing is crucial to obtain the mass of the dilaton (i.e. the conformal anomaly) and, hence, it is crucial to understand better the equivalence between gravity and conformal anomaly.
\end{itemize}

Now to the point. Let us try to understand the origin and the meaning of the time operator. In the theory of \cite{Cheung:2007st}, there is a very simple relation between the $\pi$ field and the comoving curvature perturbation ${\cal R}$. Indeed, we can write (see formula 1.14 of \cite{Baumann:2014nda})
\beq
{\cal R}=-H\pi+...
\eeq
where the ellipsis denote terms that are higher order in $\pi$.
As we already mentioned above, we identify $\pi$ with $\sigma$ and, therefore, the time operator is identified with the comoving curvature perturbation ${\cal R}$. Remarkably, this identification is in harmony with the differential equations we are working with. This point must be further elaborated. If we define $v\equiv y {\cal R}$, where (see equation 1.18 of \cite{Baumann:2014nda}) $y^2=2 M_p^2 \epsilon c_s^{-2}$, then the Mukhanov-Sasaki equation (which is the differential equation for ${\cal R}$) can be written as \cite{Baumann:2014nda}
\beq
{\ddot v}_k+3H {\dot v}_k+ \frac{c_s^2 k^2}{a^2}v_k=0.
\eeq
This is the equation of a simple harmonic oscillator with a friction term. Let us check explicitly whether this equation is in harmony with our analysis based on coherent states. In a low-redshift universe 1) the UV fluctuations of $\zeta$ are integrated out in a low-redshift universe and, hence, $H$ is basically constant; 2) The background dilaton is basically $v_k$ (with small $k$) and therefore ${\dot v}_k$ is constant; 3) the gravitational field of the black hole is very weak and basically constant. In this situation the term $3H {\dot v}_k$ of the Mukhanov-Sasaki equation is basically constant and it represents the perturbation of a constant gravitational field to the equation of the harmonic oscillator. 

Summarizing, the time operator coincides with the comoving curvature perturbation and, hence, its nature is not difficult to understand. Remarkably, we can describe a harmonic oscillator perturbed by the black hole exploiting the Mukhanov-Sasaki equation but this is possible only in a low-redshift universe where the gravitational field of the black hole is basically constant. In this sense, our equation for the coherent states is an {\it extended Mukhanov-Sasaki equation}.

\section{The physics of time}
\subsection{Toroidal quantum gravity and 2-time physics}
\label{2T}

\subsubsection{A toroidal effective theory}

In \cite{Cohen:1998zx} the authors analyze the cosmological constant problem exploiting not only the UV cut off but also an IR cut off.
The main formula of \cite{Cohen:1998zx} is
$(\Lambda_{UV})^4 < (M_p^2)/(L^2)$
where $\Lambda_{UV}$ is the UV cut off, $M_p$ is the Planck mass and L is the length related to the IR cut off.

Let us come back to our model near the big bang: here the orbifold size is not small (it is intermediate) and the string mass is large. We choose the string mass as UV cutoff and the interbrane distance $\rho$ as the (inverse) IR cut off.
Hence, near the big bang, we have
$\Lambda_{UV}=M_S$ and $\Lambda_{IR}=1/L=1/\rho$.
Furthermore, if we write the Dark Energy density as $\Lambda_{UV}^4$, the formula by \cite{Cohen:1998zx} gives us a meaningless result in our model, namely a Dark Energy density comparable to
$\Lambda_{IR}^4 exp(-4 \zeta \sigma).$

This is not what we have in our model. However, in our model the cut-offs are both dynamical: the string mass is decreasing with time after the big bang \cite{Zanzi:2012ha} and the interbrane distance is decreasing with time after the big bang. Hence, during the post big bang cosmological evolution the UV and IR cut-offs must be exchanged with each other.

We use the formulas written near the big bang and we perform the exchange of the cut-offs by hand. Hence, now, the formula obtained from \cite{Cohen:1998zx} gives a reasonable dark energy density, namely
$\rho_{DE} \simeq \Lambda_{UV}^4 exp(-4\zeta \sigma).$

As far as time is concerned, we can use our effective theory in 2D already mentioned above, where the two dimensions are (1) the interbrane distance $\rho$ (i.e. the radion) and (2) a length comparable to $V^{1/6}$, namely the dilaton ($V$ is the volume of the 6 extradimensions different from the orbifolded one). In this way, near the big bang, we have one small cycle of the torus (corresponding to the UV cut off) and an intermediate cycle of the torus (corresponding to the IR cut off). When we say ''near the big bang'' we mean that the dilatonic time is small but this does not prevent the existence of an asymptotic observer located at a large (timelike, as we will see) {\it radionic} distance from the black hole. The cosmological evolution of 13 Gy after the big bang can be summarized with a modular transformation of type S for the $T^2$ torus. The two cycles must be exchanged with each other and, hence, the dilaton must be exchanged with the radion. 

We are led to an interesting scenario. For an observer near the black hole (in strong coupling), the interbrane distance is intermediate and the volume of the remaining 6 extradimensions is small. We can connect this observer to a low redshift asymptotic observer exploiting the relativity of time. At low redshift, the interbrane distance is small and the centre of mass is located at a large timelike distance from the black hole. However, if we treat space and time on equal footing, we can also build another asymptotic observer separated by a large {\it spacelike} distance from the black hole. For this new asymptotic observer, time is fixed at the instant of the big bang while space is decompactified in some dimension and the radion has large spacelike variations.

This scenario is interesting but a spacelike radionic dimension has a number of drawbacks:\\
1) The kinetic term of the dilaton in the S-frame has the wrong sign. This might be related to the presence of extra timelike dimensions (see footnote 3, p. 404 of reference \cite{Ortin:2015hya}).\\
2) If the 4 dimensions of the brane are all spacelike, why we do not see a 4th spacelike dimension?\\

For these reasons {\it we suggest that the dilaton and the radion are both timelike coordinates in this model}. Hence, we are led to a {\it torus of time}.
Interestingly, however, when we will quantize gravity the difference between spacelike and timelike will not be physically relevant anymore.

The geometry of this effective theory is the same at small redshift and near the big bang.
The theory is the same at very large and very small redshift. To see this however we must (A) exchange the UV with the IR cut-off,
(B) exchange the dilaton with the radion.

There are a number of consequences of this approach: 1) our model is a model of holographic DE in the sense of \cite{Cohen:1998zx}. 2) When we consider a circle of dilatonic time, the $\mathbb{Z}_2$ parity of the orbifold is a ''vertical map'' (namely a map pre-post big bang), while the modular transformation of S type is a ''horizontal map'' namely a high-small redshift map valid not only in the post big bang but also in the pre big bang. Consequently, the physics does not change under $\mathbb{Z}_2$ parity, modular S transformations and compositions of the two. 3) The vacuum energy is small today but it is small also near the big bang (in the deep UV region) and this is good for structure formation. 4) At small redshift the theory is basically dilatonic (the radion is integrated out), but near the big bang the theory is basically radionic (and the dilaton is integrated out). The dilaton is stabilized at high redshift just as the radion is stabilized now. Both fields are chameleonic. 5) The symmetry under S-transformation is a symmetry high-small redshift and remarkably the presence of two orbifold fixed points is in harmony with the S-transformation. 6) The vacuum energy is small near the big bang (diluted in a big timelike radionic dimension), becomes large at intermediate stages (where the theory is interacting) and then becomes small once again at small redshift (diluted in a big timelike dilatonic dimension).

\subsubsection{Global SUSY and the cosmological constant}

Coherent states are telling us to consider ''dilatonic signals'' traveling along the orbifold of time. The dilaton participates into the description of gravity, hence, these are basically ''gravitational signals''. Actually, there is more than this: as we will see, the dilaton (and the radion) are the fundamental quanta for the description of quantum gravity. When a dilatonic signal from the brane reaches the black hole, it will not come back to the brane exactly on the same path: the path will be slightly different because near the big bang the radion is dynamical and hence a small radionic shift will select a totally different dilatonic time dimension for the way back to the IR region. A slightly different radial path in the UV means a totally different radial path in the IR. Consequently, a gravitational signal emitted from the IR brane in a point A at (dilatonic) time -t travels radially to the black hole and it is reflected (exploiting the orbifold of time) by the black hole at (dilatonic) time t=0; then it comes back radially to the IR brane hitting the brane in a point B at (dilatonic) time +t. A and B have the same dilatonic time coordinate and (almost) the same radionic time coordinate (in harmony with the fact that, in the IR, the radion is integrated out and the interbrane distance is almost zero). Hence, in the IR, A and B are separated from each other by a large {\it spacelike} distance. The two radial paths form a non-vanishing angle parametrized by a shift in the value of the radionic time near the big bang. A and B can be considered as two points on the same S-brane. 

Remarkably, with this mechanism, as far as all practical purposes are concerned, gravity propagates on the S-brane from A to B at an almost infinite speed. In other words we can exploit a non-minimal coupling term on the brane (as already mentioned above) but this theory is not able to describe completely the gravitational interaction. Indeed, new phenomena take place, they are due to the chameleonic behaviour of the time coordinates and to the orbifold of time, they cannot be described at this stage by a field theory on the brane and, from the standpoint of physics on the S-brane, they resemble the action-at-a-distance of Newtonian gravity.

In a future line of development, the presence of global SUSY in weak coupling must be analyzed in connection to \cite{Witten:2012bh} and exploiting also localization techniques in QFT (see e.g. \cite{Hori:2003ic}). The potential presence of $N=2$ global SUSY in the IR should be discussed in connection to symmetry enhancement phenomena.

\subsubsection{Brane spacetime as a shadow}
\label{gfix}

In 1909 Minkowski published ''Space and Time''. This is the majestic work containing the famous sentence ''space for itself, and time for itself shall completely reduce to a mere shadow, and only some sort of union of the two shall preserve independence''. This sentence is a crucial achievement and the modern theories of fundamental physics are based on this sentence. For example the Standard Model is a QFT formulated in Minkowski spacetime. Another example is provided by GR where gravity is described exploiting the concept of spacetime.

These considerations are certainly true but, as far as gravity is concerned, in our model we do not go beyond the semiclassical level in the effective action. Our intention here is to describe gravity on the brane at the quantum level. What are the relevant degrees of freedom in our M-theory model of quantum gravity? To answer this question we can exploit the effective action of our chameleonic model. The chameleonic equivalence principle tells us that quantum gravity is equivalent to a conformal anomaly. The universe near the big bang where the dilatonic time dimension is small (like in the Horava-Witten gravity) can be linked to our low redshift universe exploiting a conformal transformation parametrized by the dilaton. In other words, we want to link the dilatonic time dimension near the big bang with the dilatonic time dimension at low redshift and the link is provided by the conformal transformation. Remarkably, this rescaling is not global because the distance from the black hole is not constant and hence the relativity of time is not correctly taken into account with a constant rescaling of the time dimension. We infer that our transformation is a local conformal transformation. Naturally, this transformation acts also on time and, therefore, it acts on the torus of time. 

This picture is taking shape: the relevant group to describe quantum gravity is the {\it 4D local conformal group} of the MFM. The careful reader might think that this is a bit too fast because in the lagrangian of the MFM we have simply a global scale invariance, not a local one. As far as this point is concerned, let us remember that in the IR the theory is unitary, consequently, local conformal invariance can be safely assumed (see e.g. \cite{Dorigoni:2009ra}). One problem might arise when we include gravity, however, the metric is only a redundant description of gravity in our model: the chronons are the fundamental quanta of Nature. This point requires a more detailed discussion and we will proceed stepwise reviewing the reasons why our M-theory model is based only on chronons. To be more specific, our purpose is to show that gravitons and the particles of the SM are not relevant degrees of freedom in our quantum gravity model.  \\

{\bf Gravitons:}\\

We want to show that in our QG model the metric is a by-product of the chronons. We start pointing out that, from a string theory perspective, the graviton is a closed string. In M-theory, on the other hand, we have a theory of extended objects. How can we interpret the two times in a M-theory language? The dilatonic orbifold (and also the radionic curve) is simply a string, namely, it is a wrapped brane of M-theory. This {\it string of time} is made of chronons piled up together to form a timelike dimension and, hence, {\it the chronons are the building blocks of the strings}. In other words, the chronons are a (transplanckian) UV completion of string theory. Consequently, the graviton of string theory is simply made of chronons in our model and it is not the fundamental degree of freedom of the theory. The fundamental d.o.f. of our M-theory model is the chronon. For this reason we do not worry about the metric when we deal with global/local scale invariance. However, things are not so easy. Indeed, when we compose two angular momenta $j_1$ and $j_2$ in QM to obtain a total angular momentum $j$, we know that in the case $j_1=j_2=0$ we must find $j=0$. Therefore, if the dilaton is a scalar, how can we obtain a spin-2 object? At this stage, the best answer is that in this model, as we will see, the concept of spin is not physically relevant anymore. We will come back to this issue also in the Conclusions. \\

{\bf Standard Model particles:}\\

Here the trick is to interpret the standard matter fields that we use to describe physics locally (for example, the electron) as redundancies. In this way the chameleonic behaviour of the dilaton (and of the radion) is a direct consequence of a gauge fixing procedure for quantum gravity. This point must be further elaborated. The situation is completely analogous to QED. In that case we have a local symmetry of the form
\be
A_\mu'=A_\mu +\partial_\mu \lambda
\ee
and we can use this symmetry to fix the gauge. Let us consider for example the Coulomb gauge. We can start with a field $A_\mu$ where $\nabla \cdot {\bf A}\neq 0$ and we can obtain a gauge transformed field $A_\mu'$ satisfying $\nabla \cdot {\bf A'}= 0$. To obtain this result we must choose $\lambda$ so that
\be
\nabla^2 \lambda=-\nabla \cdot {\bf A}.
\label{cam}
\ee

Now let us come back to quantum gravity. The gauge transformation is a local conformal transformation and we can use it to fix the gauge. The gauge parameters are the dilaton and the radion. Matter fields (and also the gauge fields of the Standard Model) are coupled to the chameleons and hence, they play the role of $A_\mu$ in our QED example mentioned above. When we fix the gauge we create a link between the gauge parameters and the non-gauge-fixed fields (see equation \ref{cam} in QED): this is precisely the origin of the chameleon mechanism.  Hence the components of the matter fields and also the gauge fields of the Standard Model are simply redundancies and when we fix the gauge these degrees of freedom are not physical anymore. The only relevant degrees of freedom are the gauge parameters, namely the dilaton and the radion. \\

However, a word of caution is necessary. It would be necessary to analyze the role of the conformal group exploiting the full SUGRA theory in eleven dimensions. Our approach here is simply to exploit a 4D effective lagrangian and we analyze the role of the conformal group simply exploiting this 4-dimensional point of view. Another related problem is that we should link the conformal anomaly studied by Fujii with the existing literature on conformal anomalies. These issues are left for future research.

We infer that the chronon (namely the time) remains as the only physical degree of freedom for quantum gravity on the brane. However, this is true only in a specific ground state related to the gauge fixing procedure. In other words, our theory provides a no-go theorem for quantum gravity.\\

{\bf No-go theorem for quantum gravity:}\\
An observer cannot analyze quantum gravity data in his laboratory (i.e. in his local ground state) because when he tries to quantize gravity the gauge fixing procedure makes the SM unphysical and the observer himself is not physical anymore with respect to quantum gravity data. However, an observer can (at least in principle) analyze quantum gravity in a ground state different from the ground state where he lives.

$\Box$

 Therefore, for example, an observer living in the ground state of Bern can analyze quantum gravity on cosmological scales because the far IR region corresponds to a ground state which is not the ground state of Bern and, hence, the observer is physical because the SM particles in Bern are not gauged away.

As far as space is concerned, we must quantize gravity to clarify this issue. As we will see in paragraph \ref{chrononsize}, QG can change a spacelike interval into a timelike one. Hence, space is once again described by chronons. If we prefer to imagine that our QG theory is purely spacelike, this is possible.

Summarizing, the theory is telling us that, whenever we consider one specific ground state on the brane, Nature is described at a fundamental level only by one quantum: the chronon. There is only one fundamental interaction, namely gravity. The equation governing the dynamics of the chronon is the (modified) Schroedinger equation and, hence, at this stage, the Schroedinger equation is the most fundamental equation of physics for a specific ground state. Spacetime on the brane is a shadow and only the chronon is the physical degree of freedom at a fundamental level in a specific ground state on the brane. All the fields that we use to describe standard physics (e.g. the electron, quarks...) are simply redundancies in a specific ground state and all their components are gauged-away by the generators of the 4D local conformal group. 

Two comments are in order:

1) Remarkably, the torus of time is not continuous but it is simply a lattice\footnote{Actually, as we will see, it will not be a Bravais lattice but it will be more similar to an amorphous solid. At this stage we use the word ''lattice'' to emphasize the non-continuous nature of the structure.}. In a more geometric language, we have a fibre bundle: the torus of time can be seen as a base manifold while space plays the role of the fiber. The global scale invariance of the MFM is a remnant global symmetry after the gauge fixing in 4D and this is completely analogous to the conservation of electric charge in QED (related to a global U(1) symmetry) which is guaranteed even if we fix the Coulomb gauge.

2) We are {\it not} saying that Einstein was wrong. Our theory is relativistic and it can be described in terms of metric (at a classical or semiclassical level) when we work in 11 dimensions. Naturally, this is a very peculiar spacetime with two timelike dimensions, but it is an 11D physical spacetime. However, if our intention is to build an effective theory of {\it quantum} gravity {\it on the brane}, then the concept of spacetime is not useful anymore and only the chronon remains as physical degree of freedom in a fixed ground state.

\subsubsection{Signals from the future (of a different dilatonic time dimension)}
\label{timetravel}

In this model, the dilaton and the radion play the role of time coordinates. If we modify the matter density of the environment, we change the value of these chameleon fields. Our effective MFM lagrangian is not describing a bi-scalar-tensor theory of gravity. Indeed, we have only one scalar field in the effective MFM lagrangian. Naturally, this single scalar field is the result of a proper redefinition of what we mean as ''dilatonic time''. After all, the definition of time is not unique and this is true already in standard Special Relativity: we are free to redefine time. After this redefinition has taken place, if we interpret the dilaton also as a local time, then our time is dilatonic and the radion is currently stabilized. Obviously, the chameleonic shift related to a variation of the matter density cannot be interpreted as a time travel. Actually, when we shift the matter density, we are creating an alternative time dimension. This is due to the chameleonic behaviour of the radion. Let us further elaborate this point.

Let us imagine a gravitational wave (GW) detector (e.g. VIRGO) in a certain dilatonic time coordinate (e.g. the year 2020). Let us suppose that our GW detector is placed inside a chamber where we can choose the air pressure in the way we prefer. Moreover, we place inside the chamber a GW generator. At the beginning of our experiment, the air pressure in the chamber is the standard one. Now we increase the air pressure and, hence, the  density of the environment is larger than before. Therefore, the dilaton is shifted to a smaller value and, in this way, we can synchronize two dilatonic clocks together: 1) the dilatonic clock of our high-density experiment in the year 2020 (let us call it ''future apparatus'') and (2) the dilatonic clock of the same experimental configuration in the year 2019 at standard density (let us call it ''past apparatus''). With this method the two experiments are, as far as the dilatonic time coordinate is concerned, chronologically aligned: the two experiments live on the same S-brane but in two different dilatonic time dimensions. In particular, in this example, we assume that the radionic time of the future apparatus is smaller than the radionic time of the past apparatus. In other words, the dilatonic coordinate is the same for the two experiments but the radionic time is (not exactly) the same because also the radion is a chameleon and the shift in the air density modifies (slightly, because the radion is stabilized) the radionic time coordinate. A small shift in the value of the radion (which is stabilized by some UV dynamics) selects a totally different dilatonic time dimension and this effect is reminiscent of the butterfly effect. 

To proceed further, let us suppose that the future apparatus generates a gravitational wave in the year 2020 (interestingly, also the GW can produce a jump of the source in another dilatonic time dimension). The CEP tells us that the dilaton is coupled to the gravitational field of the GW and, naturally, it is also coupled to the matter density of the apparatus through the chameleon mechanism. Consequently, the dilaton ''takes a picture'' of the experimental configuration (including the GW) and brings this signal through the orbifold of time along a dilatonic time dimension that we call ''number one''. When our dilatonic signal is very close to the big bang, also the radionic time becomes dynamical and this radionic evolution selects a different dilatonic path. Hence, the dilatonic signal will come back to the IR region, but not along the dilatonic time dimension ''number one''. Indeed, it will follow another dilatonic time dimension that we call ''number two''. This means that our dilatonic ''picture'' of the future apparatus will come back along a different dilatonic time dimension exploiting the radionic time evolution. The dilatonic signal will explore various possible dilatonic time dimensions (indeed the torus of time is discrete, not continuous) until it will enter into the dilatonic time dimension where the past apparatus is located.

If the characteristics of the environment of the two experiments are precisely the same, then the environment of the past apparatus is a ''hook'' for the signal: the dilatonic signal will reproduce our GW in the past apparatus where our GW detector is placed and the presence of the GW will slightly modify the value of the radion putting our laboratory in a different dilatonic time dimension (where e.g. wonderful research results might be discovered immediately exploiting the data received through the GW signal). Let us further elaborate this point. When the dilatonic signal reaches the dilatonic time of the past apparatus (in the correct dilatonic time dimension), the entire fiber is ''available'' to the signal. Indeed, both time coordinates are completely specified and the signal is placed in a specific point of the torus of time. However, one specific point on the torus of time corresponds to an entire fiber of space and the dilaton can enter into the torus (i.e. the base manifold) starting from an arbitrary point on the fiber. Equivalently, we can say that the dilatonic signal reaches simultaneously all the particles/fields sharing the same time coordinates (and this is in harmony with the fact that the dilaton parametrizes the cosmological expansion of the entire universe, not only of a part of it). The signal has two different components: 1) an environment-related component which is static and (2) the GW signal which is oscillating. The environment-related component is compatible only with the experimental set up and provides the hook to receive the message. Indeed, if the signal is localized in a point on the brane different from our experimental site, then the static component of the signal would lead to a static gravitational field which would violate energy conservation, because we do not have only the signal at the past apparatus but also other gravitational fields in the entire fiber. Therefore, if we believe in energy conservation, we infer that the signal must be received by our past apparatus\footnote{In principle we might wonder whether there is a ''copy'' of the experimental apparatus somewhere on the fiber, but we exclude this possibility at this stage.}. When the dilatonic signal reaches the past apparatus, all the environment-related characteristics of the signal produced by the future apparatus fit perfectly with the environment of the past apparatus, while the part of the signal generated from the GW generator of the future apparatus will be reproduced and detected by the GW detector of the past apparatus. The signal traveled from the year 2020 (to the big bang and back in a different dilatonic time dimension) to the year 2019. \\

{\bf Discussion:}\\

The careful reader might be puzzled by our considerations, in particular: 
\begin{itemize}
\item The reader might wonder whether the dilatonic signal becomes distorted as it propagates. Happily, the two experiments have almost the same time coordinates. Indeed the dilatonic time is the same, and the radionic is shifted only slightly. Therefore, even if the signal can be distorted along its travel on the orbifold, no distortion is expected in the signal received by the past apparatus. 
\item The reader might say that our future is not yet physical because we have not yet lived it, hence we cannot receive signals from the future. The answer to this problem is that the emission of the signal would take place in the future {\it of a different dilatonic time dimension}. Indeed, the cosmological model is cyclic and, consequently, in principle, we can have plenty of dilatonic time dimensions that have already been ''lived''. Among these dimensions, among these Universes of the past cycles, we might find also worlds which are almost identical to ours, but these worlds have already evolved in time! Naturally, the probability that, in one of the past cycles of the universe, a GW signal has been emitted from a world almost identical to ours is related to the number of these past cycles. The larger is the number of past cycles, the larger is, in principle, the aforementioned probability. There is also another comment regarding the existence (in another dilatonic time dimension) of a world almost identical to ours. Indeed, if we believe in Everett-DeWitt's Many Worlds Interpretation of QM, we can easily imagine worlds which are almost identical to ours. To be more concrete we can imagine a Stern-Gerlach apparatus and we prepare our initial ket in a linear combination of up and down spin. In our world the result of the measurement is, e.g., ''spin up'', but also the ''spin-down'' result is physical (in another world), the difference between spin up and down is felt by the chameleonic radion and this produces a shift in the dilatonic time dimension. 
\item One might wonder why a signal should be emitted from a civilization. Every civilization wants to leave a heritage in the future. If we imagine that Nature has chosen this theory and if we imagine an evolved civilization aware of this fact, then it is plausible that this civilization wants to prepare a transfer-of-knowledge from their cycle of the cosmological evolution to another cycle. In this case, the cultural legacy of the civilization of the past would realize this past civilization in our world and the idea of human evolution should be discussed in the framework of cyclic cosmology. 
\item We introduced an IR cut off related to the orbifold size. However, how can we be sure that the size of the orbifold is finite? After all, the size of a timelike dimension is observer-dependent when we take into account the relativity of time. The answer to this problem is related to the presence of the bulk dilaton in 5D. If we imagine an infinite orbifold, then we should remove the stack of branes and we would work only with the hidden brane. Consequently, the number of moduli of the SUGRA action would be one and not two. 
\item One relevant aspect of our analysis is causality. Indeed, when the dilatonic perturbation flows towards the big bang along the orbifold of time, we might think that the perturbation is simply lost because we ''play the film backwards''. How can we reconcile causality with this propagation backward in time? The simplest possibility is to exploit S-duality. Let us illustrate this point. In our everyday life physics is causal and the arrow of time is perfectly well-defined. Signals travel from the past to the future. However, S-duality identifies the weak coupling with the strong one and, hence, at the end of the cosmological expansion (and this expansion must finish because the bulk dilaton is present in the model), our dilatonic timelike line forms an $S^1$ space which will be orbifolded exploiting the $\mathbb{Z}_2$ parity. A line connecting the two fixed points of the orbifold is already a closed loop. When the orbifold of time is formed, we obtain a structure where the arrow of time is non-physical, hence, string dualities and orbifold parity lead us to a description of physics which is effectively non-causal. These stringy effects can be summarized in an intuitive way through a cyclic cosmology. The pre-big-bang phase is simply induced by S-duality. In this way, our local physics is always perfectly causal, but string/M-theory tells us that some stringy effects can be conveniently summarized through non-causal petals of the flower of time (see the Introduction). It is for this reason that the signal is not lost when it comes back to the big bang along the dilatonic timelike dimension. Causality is a concept related to the manifold of time: locally it is a good guideline, but globally (i.e. when we consider at least an entire petal) it is effectively lost. 
\item What happens at the end of the flower of time? Do we overwrite with new physics the existing petals or we simply repeat eternally the same events? This point will be further discussed later. At this stage we simply add a few qualitative comments. At the end of the last petal, locally, causality is a good guideline. Hence, gravitational perturbations from the past petals do not allow a simple repetition of the flower. This is a very good news because the chances of evolution of the mankind are higher in this way. A new dimension is necessary to avoid causality violations. Consequently, the correct theoretical framework seems to be F-theory \cite{Vafa:1996xn, Morrison:1996na, Morrison:1996pp} (for an introduction see \cite{Denef:2008wq, Weigand:2018cod} and references therein) and we obtain a collection of flowers forming a cylinder where the new dimension is the symmetry axis of the cylinder. Hence, our M-theory model seems to be ''embedded'' into F-theory. It would be interesting to understand whether $k_{11}$ of our M-theory model is related to the size of the 12th dimension of F-theory. From the mathematical point of view, we can simply say that (dimensionless) time is a complex number $z$, a flower of time is an annulus in the complex plane and the various flowers of time are reminiscent of Riemann sheets of a multivalued function $f(z)$. One more comment is necessary. A countable (and not dense in the reals, more on this in paragraph \ref{6G}) infinity of chronons along the radionic Closed Timelike Curve (CTC) seems to imply a ''forbidden chronon''. Indeed exploiting 1-point compactification, the size of the chronons would go to zero at the North pole of the $S^1$ manifold. Since we do not believe in pointlike particles (this is a personal point of view of course, see paragraph \ref{6G}), the orbifold at the North Pole of the CTC seems to be forbidden (we will further discuss these issues in paragraph \ref{chrononsize}). This might be the point where we change the Riemann sheet. Interestingly, the flower of time is not isotropic because gravity is not only related to the dilaton (i.e. the CEP) but also to the radion. 
\item In this model, we are exploiting the accelerated fall of the chronon towards the black hole in the pre big bang phase to explain inflation in the post big bang phase. One might say that this is not correct because during our cosmological expansion we have not yet formed the orbifold of time (which is a global structure). This point can be clarified, at least partially, if we exchange the UV region (strong coupling) with the IR (weak coupling) exploiting S-duality. Indeed, if we do in this way, the standard backreaction effects (i.e. effects from the local physics to the global one) are reinterpreted as effects from global to local. Consequently, the role of local-to-global effects to analyze the Dark Energy problem is S-dual to the role of global-to-local effects to analyze inflation. For a discussion of backreaction effects in this model the reader is referred to \cite{Zanzi:2010rs}. 
\item One interesting issue is related to the lightlike nature of the horizon. How can we claim that the branes are S-branes and, nevertheless, they play the role of a horizon? The answer is that we have two times and, hence, even if the horizon is ''orthogonal'' to the dilatonic time (and this leads us to the concept of S-brane), it is not orthogonal to the radionic time and the fluctuations of the radion still provide a timelike ''tick'' (in this sense we can define a lightlike horizon). 
\item Another interesting issue is related to the assumed presence of the stack of branes. How can we trust this model if we neglect a large number of moduli in the effective action? Indeed, we have many branes inside the stack but we introduce only one modulus to parametrize the position of the stack. The answer is that the position of the branes inside the stack is related to the matter density: in this model the air of a room lives on a certain S-brane, but the wall of a room is localized in a different S-brane, whose coordinate in the fifth dimension is different. Hence, our universe is the result of the ''superposition'' of different parallel universes and, in this sense, the choice of a single (but chameleonic!) modulus is reasonable.
\item It is common knowledge that
EM waves travel in vacuum with speed $c$. Naturally, they can be quantized and the related particles (i.e. the photons) travel in vacuum with the same speed $c$. The same is true for GW (see \cite{Monitor:2017mdv}). Since a GW travels in vacuum with speed $c$, then it should be quantized in terms of massless particles and not massive particles. However, the dilaton (and the radion) can be massive and, therefore, they should travel with speed $v<c$. This is not a problem because the dilaton, namely the physical time in our post big bang universe, is massive granted that the matter density of the environment is not small. Therefore, when we consider the propagation {\it in vacuum} of EM waves or GW, then the expected propagation speed in the MFM is $c$, in harmony with the experiments. However, when we consider the propagation of EM waves or GW in a high matter density medium, then the propagation speed in the MFM should be smaller than $c$ for both waves. Now, let us consider a photon. It is a massless particle which moves in vacuum with speed $c$. However, when we consider the motion of a photon inside the Sun, we know that the photon will interact with matter and, hence, the time necessary to travel from the center of the Sun to the surface of the Sun is not given by $R_\odot /c$ (where $R_\odot $ is the radius of the Sun). The photon is slowed down by the interaction with matter and this is true in standard theories (not only in the MFM). In this sense, the photon acquires an effective mass. This is basically a chameleonic behaviour of the photon but naturally we are not saying that the photon acquires a longitudinal polarization state. The number of d.o.f. should be counted in the far IR (i.e. in vacuum) and then the matter of the environment can produce an effective mass for the photon. In general we expect that also a GW slows down in matter, but this effect in general should be smaller than an EM wave in the same medium. Hence, the next problem is to recover the various interactions (including the SM) starting from the chronons. This issue will be reviewed, at least partially, in the next section.

\end{itemize}

\subsubsection{The Standard Model and the chronon}

If our intention is to describe quantum gravitation exploiting the dilaton and the radion, this means that in the deep UV region (at energies higher than the string mass near the big bang), these two particles are the fundamental degrees of freedom (d.o.f.). Hence, when we move to the IR, we should be able to ''reconstruct'' the quantum SM starting from the two fundamental d.o.f.. In other words, is it possible to reconstruct the SM starting from the dilaton and the radion? The reader might think that this is not necessary because in our model the SM is redundant. However, the redundancy of the SM is true only locally (i.e. in a specific ground state where we quantize gravity) and not globally. We will discuss spin, mass and interactions.\\

{\it Spin and mass:}

First of all, in our model, there are quantum gravitational effects on the brane which are reminiscent of Newtonian gravity. Consequently, the concepts and symmetry groups typically used in relativity are not important anymore when we analyze quantum gravity {\it on the brane}. For these reasons, on the brane, the Lorentz group is not expected to play an important role anymore. It is common knowledge that the Lorentz group is crucial in QFT and, in particular, it leads to the correct description of the concept of spin. Therefore, we expect that the concept of spin becomes useless when we analyze our QG theory on the brane. Let us further elaborate this point searching for some ''evidence'' of this surprising spin-independence. Is the chameleonic behaviour spin-independent in the MFM? The dilaton and the radion are scalar fields and they are chameleons. Matter fields are chameleons in this model because the Planck mass is renormalized. The photon acquires an effective mass inside the Sun, as already mentioned above and this is basically a chameleonic behaviour of the photon.\\

{\it Interactions:}

If our intention is to reconstruct the SM starting from the dilaton and the radion (which are gauge singlets), then it would be necessary to show that the gauge groups and their representations are not physically relevant anymore. In this case, indeed, the difference between a photon and a gluon would disappear, the difference between a left-handed electron (a doublet of $SU(2)_L$) and a right-handed electron (a singlet of $SU(2)_L$) would disappear and all the elementary particles of the SM would be considered as gauge singlets (obtained from our UV chronons). 

How can we remove the physical importance of a gauge group? Naturally, the relevant gauge group is related to the length scale of the problem we are considering. For example, there is a general (and well-known) phenomenon: the symmetry groups of atoms are not the symmetry groups of molecules and these are not the relevant symmetry groups of crystals. This idea tells us that we can remove the physical importance of a gauge group by changing the length scale of the problem (i.e. by considering an RG running). Remarkably, the CEP tells us that the RG-running is due to the conformal anomaly, namely it is a quantum gravity effect. Quantum gravity can select or discard a symmetry group and, in this sense, there is no difference between a left-handed electron and a right handed electron from the standpoint of the chronon. Let us analyze this point in a more quantitative way. The MFM lagrangian is used in the UV but also in the IR (we have a unique lagrangian valid everywhere). The RG running (i.e. the time evolution in this model) is summarized by a dilatonic shift $\Delta \sigma$ and, remarkably, this shift does {\it not} modify the representation of the gauge group. Let us discuss the consequences of this fact. In the MFM we used a scalar $\Phi$ as representative of matter fields. Now we see that this choice can be interpreted as a natural consequence of the spin-independence of the QG model. The choice of a scalar field for matter simplifies the analysis of the breaking of scale symmetry, but this choice is not less physical than the fermionic one in our QG model! Hence, in the far IR, where global SUSY is unbroken, the $\Phi$ field is representative of the entire chiral superfield of matter. In the far IR, the cosmological matter is neutral (the charge is globally zero) and when we run to the UV exploiting a dilatonic shift, the representation of the $\Phi$ (super)field is not modified. Nevertheless, we keep on using the $\Phi$ field as representative of microscopic matter (e.g. matter fields of the SM, which are obviously electrically charged). Hence, the representation of the gauge groups where we put the fields is not interesting anymore. In this sense, the SM can be obtained starting from the chronon.

Now we can come back to the problem mentioned above: why the propagation speed of a GW in a high density medium is reduced less than the corresponding speed of an EM wave? Naturally, the point is that the photon interacts also electromagnetically while the GW interacts only gravitationally. From the standpoint of the MFM, this extra-interaction is simply parametrized by a proper dilatonic shift $\Delta \sigma$, because in this way we can introduce the electromagnetic U(1) gauge group. Consequently, the difference in the propagation speed of an EM wave and a GW in a high density medium is parametrized, in the MFM, by a QG effect. Once again, only the gauge singlet chronon is physically relevant in our QG model.

One comment is in order. It is common knowledge that a selectron has not been detected experimentally. The standard answer is that global SUSY (if present in Nature) must be spontaneously broken at a proper energy scale. In this way, the mass of the selectron will be large enough to escape detection. In our model, the interpretation is different. It is true that we have global SUSY, however, we cannot forget about gravity and this interaction at this stage is neither described by Einstein's theory nor by SUGRA. Indeed, as already mentioned above, we have global SUSY (globally restored and locally broken) plus a sort of Newtonian gravity on the brane. Therefore, at LHC, global SUSY is broken because the vacuum energy is positive (the chameleon is coupled to local fields), but the standard mass splitting between (e.g.) an electron and its superpartner is absent: quantum gravity effects produce a surprising spin-independence and also a surprising representation-independence. Therefore, the electron is representative of the entire chiral superfield even if global SUSY is broken at LHC. Once QG effects are taken into account, there is no physical difference between an electron and its superpartner and, hence, the mechanisms which give a mass to the fields are precisely identical. In this sense, our model is telling us that we have already found supersymmetry in Nature: the standard particles are the only representatives of the superfields (once quantum gravity corrections are taken into account).

Let us add some more comments about the redundancy of the SM.

As already mentioned above global-to-local effects are useful to understand inflation. Indeed, the accelerated motion of the chronon towards the black hole in the pre big bang can explain inflation in the post big bang. If global-to-local effects are useful to understand inflation (namely an acceleration which characterizes a specific - local - period of the cosmological expansion), then they should be useful to understand something about physics in our standard laboratories. Let us try to explain the redundancy of the SM using this argument.

We start considering a standard cosmological scenario where the SM describes particle physics and the relevant time is dilatonic. The universe is expanding and we live in the post big bang phase. At the end of the cosmological expansion S-duality forms the $S^1$ manifold. The upper semicircle can be identified with a pre big bang phase, while the lower semicircle can be identified with the post big bang phase. The SM lives in the lower semicircle. However, on global scales we can use the $\mathbb{Z}_2$ parity of the orbifold to map the SM from the lower semicircle to the upper semicircle. This mapping corresponds to the $\mathbb{Z}_2$  operation $0+1=1$. Since {\it locally} we believe in the causality principle, the upper semicircle is non-physical (because it is formed only at the end of our cosmological expansion exploiting S-duality) and, hence, the SM is non-physical. The situation is analogous to a standard mirror in our everyday life. When we look into the mirror, we see a non-physical ''world'' but we can learn something useful about our physical real world. The pre big bang is like an image in a mirror: from the standpoint of local physics the pre big bang is non-physical, but it can tell us something useful about our physical post big bang phase. Enough for the $0+1=1$ step. Things get more interesting when we remember that in the $\mathbb{Z}_2$ algebra, we can write $1+1=0$. This is our next step. This means that we mix together the real post big bang dilatonic time with the non-physical pre big bang phase containing the SM. In other words, the non-physical SM in the pre big bang phase is joined together with the physical dilatonic time of the post big bang phase. The effective result is that we have a redundant SM mixed with a physical dilatonic time. This is precisely the scenario we obtained exploiting the chameleon mechanism and the gauge fixing procedure of paragraph \ref{gfix}. Remarkably, the localization of the SM in the bulk (see \cite{Zanzi:2016thx}) is surprisingly natural when we use this construction.

\subsubsection{Cosmological constant: further remarks}
\label{6G}

This model has been exploited in these years to keep under control the cosmological constant. Consequently, the reader might wonder whether the results of the present article are in harmony with the solution discussed in \cite{Zanzi:2010rs}. It is true that in this model we have a 4D lagrangian with a density-dependent amount of scale invariance \cite{Zanzi:2010rs}. It is true that this lagrangian can be (basically) obtained from a multi-brane set up of M-theory \cite{Zanzi:2016thx}. However, the description of quantum gravity on the brane is surprising: we find new gravitational phenomena which resemble a Newtonian action-at-a-distance on the brane. These Newton-like phenomena on the brane are clashing, at this stage, with a field theory (and lagrangian) description of QG on the brane and, in a sense, they are telling us at this stage that our theoretical tools are not enough to deal with the cosmological constant problem. Hence, the reader may ask: is this a solution to the cosmological constant problem or not? 
 
We can try to do more with two comments about the cosmological constant. Before we discuss this issue, it is necessary to mention a ''Personal Ansatz''. Indeed, we do not believe in pointlike objects. We are well aware that some readers will disagree about this personal point of view and, therefore, we deal with this problem by simply {\it assuming} the absence of pointlike particles or, to be more precise, we assume that the size of an elementary particle is bounded from below. If we introduce this ''Personal Ansatz'' then we are ready for the comments regarding the cosmological constant problem.

Now the comments.

1) The action-at-a-distance that we have in the IR region on the brane is mysterious, because at this stage it is not clear how a brane observer could describe it and, moreover, we do not know if (and how) it can contribute to the cosmological constant. If we use our ''Personal Ansatz'', a pointlike chronon must be excluded. Consequently, if we link our chronons to a numerical set, the chronons (representing the torus of time) are not dense in the reals and, therefore, the concept of {\it contiguous chronons} is meaningful. 

Now, in order to understand the action-at-a-distance, let us exploit QG. As we will see, spacelike and timelike intervals are equivalent in our QG theory. Consequently, this action-at-a-distance is not surprising in QG. Moreover, if we use S-duality we can map this action-at-a-distance into a UV problem and we know that in the deep UV, A) physics is classical because quantum fluctuations of chronons cannot take place without a background (actually this is an assumption mentioned in the introduction) and B) two {\it contiguous} chronons cannot interact because a UV completion of the chronon theory is totally absent at this stage.
Hence, the action-at-a-distance is mapped into a trivial contribution to the cosmological constant.

2) The second comment is related to the UV cut off of the theory. When we built our 4D effective action, the UV cut off of the theory was the string mass, which is related to the radion and it is chameleonic. However, when we include QG effects, what is the UV cut off for the chronon theory? The chronon size would be the natural UV cut off. Now to the point. What is the size of the chronon? 
To illustrate this point, let us consider the dilatonic timelike dimension. Its size is finite because we have an IR cut-off. Needless to say, we perceive our dilatonic time as continuous, but in our model this timelike dimension is the result of a large number of chronons forming a ''string''. This string starts from the black hole at the big bang and it goes on until the second orbifold fixed point is reached. That point corresponds to the end of the cosmological evolution and, in that point, exploiting S-duality, we are led to 1-point compactification (see, e.g. p. 84 of reference \cite{Nakahara:2003np}). This means that we close the string of chronons (we make the inverse of a stereographic projection) identifying the weak coupling with the strong coupling. Had we considered an infinite number of chronons, we would have obtained a chronon size that tends to zero near the North Pole of the 1-point compactification because in the North Pole we map the point at infinity and the various chronons are squeezed by the compactification. In this way we are led to a non-isotropic $S^1$ compactification for the dilatonic time: the closer we are to the North Pole, the smaller is the chronon size. However, we have a large but {\it finite} number of chronons and the 1-point compactification is compatible with our Personal Ansatz regarding a minimum chronon size. Nevertheless, we still keep the information regarding anisotropy as valid. In this way we have big (small) chronons at large (small) distance from the North Pole (i.e. the black hole) and then we impose the $\mathbb{Z}_2$ parity. The chronons of the same size are identified with each other by $\mathbb{Z}_2$ parity and we are left with a string of chronons starting from the black hole. Remarkably, in this string, the size of the chronons is an increasing function of the distance from the black hole in harmony with the relativity of time (i.e. in harmony with the chameleonic behaviour of the dilaton). The reader might disagree because in relativity the twin who remains always far away from the black hole is the older one when he meets the twin who traveled near the black hole. The answer to this problem is that we can always exploit S-duality and, hence, we can exchange the small tick with a large tick. The important point is to obtain the anisotropy mentioned above. {\it The relativity of time is a stringy effect: it is induced by S-duality.} We found another element, together with the fundamental role played by chronons, that seems to support the possibility of a non-relativistic formulation of M-theory or, more precisely, of a formulation that does not require relativity as a basic ingredient. It might be that relativity (or a part of it - we have already seen aspects of brane gravity that are not relativistic in this model) is simply a by-product of M-theory but it is not required as a starting point. One more comment is necessary, in the deep UV, the size of the chronon (i.e. the inverse of our QG UV cut off) is smaller than the planckian string length (i.e. the inverse 4D UV cut off without a chronon interpretation): in the deep UV chronons explore transplanckian physics. As we will see later, in this transplanckian regime the relevant geometry will be given by a cone.

 Enough for the dilaton. Let us discuss the radionic chronons. The duality dilaton-radion tells us that the same physics is present in the radion CTC and, moreover, quantum gravity is related not only to the dilaton but also to the radion. We infer that the flower of time is not isotropic: when we change the angular coordinate we modify the gravitational field and, hence, the chronon size. The size of the radionic chronons near the North Pole is small and, hence, in harmony with our Personal Ansatz, we have a forbidden direction (here we change the Riemann sheet). This is the reason why the theorem regarding CTC with $S^n$ spaces with $n>1$ is not a problem: {\it we actually do not have CTC}, because the forbidden chronon forces us to change the Riemann sheet. Once again, the size of the chronons is small near the North Pole and large near the South Pole of the 1-point compactification. This is the geometrical description of the radion CTC and it is fully compatible with S-duality (the radionic weak coupling is joined with the radionic strong coupling and these two regions ''touch'' each other in the North Pole). Consequently, the UV cut off of the theory is also radion-dependent. 

Now we come back to the cosmological constant problem. Remarkably, near the radionic North Pole, the chronon size is almost zero and, hence, the UV cut off is extremely large. The reader might think that this is a complete disaster for the cosmological constant problem. Not so at all! It is certainly true that a large UV cut off leads us to a theoretical prediction for the cosmological constant which is, in principle, worse. However, once the UV cut off is extremely large, the cosmological constant problem is not different from the hierarchy problem related to the Higgs mass and this is a very good news. Naturally, the theoretical prediction must be in harmony with observational data and, for this reason, in our model we exploit scale invariance (and global SUSY). 

The final scenario is reminiscent of the Large Number Hypothesis of Dirac (see e.g. \cite{Giudice:2008bi}): any very large number in nature should be simply related to the age of the universe. This is precisely the scenario we have in our model. If we wait for a long enough time, the chronon reaches petals of the flower of time where the UV cut off is so large that the difference between the hierarchy problem of the Higgs mass and the cosmological constant problem can be safely neglected.
 
Now we come back to the question mentioned above: is this a solution to the cosmological constant problem or not? At this stage, the best answer is probably contained in the words of H. Poincare' (year 1908) which seem to characterize the entire Physics: ''Il y a seulement des problemes plus ou moins resolus''.

\subsection{The Borges' Library}
\label{Borges}

{\it ''If an eternal traveler should journey in any direction, he would find after untold centuries that the same volumes are repeated in the same disorder} - {\it which, repeated, becomes order: the Order.''}
(Jorge Luis Borges)\\

In this section we will explore the geometry of time. In this paper we worked with two timelike dimensions in a model (originally proposed in \cite{Bergshoeff:2000zn}) which embodies the main properties expected from M-theory. In principle, the model might be embedded into F-theory and, in this case, an additional dimension would be present. As already mentioned e.g. in \cite{Bars:2010zz}, depending on the application the 12th dimension of F-theory is sometimes timelike and sometimes spacelike. Here we will assume it to be timelike and this choice is guided, as we will see, by a ''glassy'' nature of time. In this way we are led to a 3-dimensional {\it time} and we can build purely timelike 3-dimensional structures. The resulting geometry of time will be related to string dualities and it will be similar to the Library of Borges \cite{Borges:libro}.

\subsubsection{The space and a 6D Library}
\label{chrononsize}
 
Before we start with a detailed discussion of 3-time physics and of the related 3D timelike structures, we must clarify the role of space.

Two remarks are in order. 

1) The gauge fixing procedure summarized by the chameleon mechanism is related to {\it fields}, but this does not imply that the space coordinates are redundant. Even if we define space through some particle physics events, we did not prove that a ''totally empty'' space does not exist. 

2) In the deep UV (namely at energies higher than the string mass near the big bang, namely at transplanckian energies) the curvature singularity becomes a coordinate singularity. This point should be further discussed. A generalized cone (i.e. our bag geometry) has a curvature singularity in the origin granted that the base manifold is not a unit d-sphere (see p.51 of \cite{Kirsten:2001wz}). Now back to the model. The singularity is a curvature singularity but only when quantum effects are important, because quantum fluctuations produce a deviation of the base manifold from a unit d-sphere and, hence, there is a curvature singularity in the origin. However, if we go really deep in the UV (at transplanckian energies) the model becomes classical because the quanta do not fluctuate without a background and, consequently, the singularity becomes a coordinate (in particular conical) singularity (because the base manifold does not deviate from a unit d-sphere). In our model, surprisingly, a coordinate singularity becomes a curvature singularity when we run to the IR starting from the deep UV. {\it Gravity is induced by quantum mechanics}.

Now, we can discuss the concept of space in our model. Up to now, the observer is located on the stack of branes. The air of a room has a different density with respect to the wall of a room and, hence, we must consider different branes. The result is that the observer is living on the stack. What happens if we use S-duality? The stack moving towards the deep IR is mapped into a transplanckian stack, localized near the tip of the cone in the deep UV (where physics is classical). The hidden brane is mapped near the dual singularity (i.e. the second orbifold fixed point). Hence, the dual theory is basically a Klebanov-Witten model \cite{Klebanov:1998hh} where the observer is described by a transplanckian stack of branes moving towards the tip of the cone. Therefore, the observer is at the center of the bag. Remarkably, when we change the matter density we shift the position of the branes and, hence, of the horizon. Remarkably, even if we consider the standard Reissner-Nordstroem (RN) black hole, when $r<r_-$ (i.e. inside the inner horizon) the radial coordinate becomes spacelike. Consequently, {\it QG changes a spacelike interval into a timelike one (and viceversa)}. Once again, in this QG (dual) theory spacelike and timelike intervals are basically equivalent to each other because the RG running, namely QG effects, are able to shift the position of the horizons of the black hole. For this reason, we do not remove space from our model and we are led to a scenario where the 5D bag is a floor of a 6D Library (more details on this later). The 6th dimension corresponds to the 12th dimension of F-theory and it is ''orthogonal'' to the bag. Hence, our cosmological model is S-dual to a ''geocentric cosmology'': the observer is at the center of the bag in the dual theory, the radial coordinate is spacelike and the $S^4$ boundary is composed of two spacelike and two timelike coordinates (the radionic time is basically fixed while the dilatonic is physical as usual). We obtained a 6-dimensional Library but the effective description at our macroscopic scales is 4-dimensional because the vertical time of F-theory is fixed on the floor where we live and the radionic time is fixed too. We are left with a 4-dimensional effective theory with Euclidean time: happily a very familiar theory that we will use in the future for phenomenological investigations.

In the next paragraphs we will ''forget'' about space and we will work with 3-time physics. Indeed, to simplify the analysis, we will consider an $S^1$ boundary on the floor of the Library as representative of the physical $S^4$. The idea is that the physical results on the timelike $S^1$ should be extended identically on the entire physical $S^4$ because QG can change a timelike interval into a spacelike one so this difference is not so important anymore. Basically we will simply factorize our formulas: for example the single sine function in \ref{2Dsbis} represents a product of sine functions with one sine factor for each coordinate of the $S^4$.

\subsubsection{3-time physics}

At this stage, in our model, (dimensionless) time can be conveniently described as a complex number $z$. The annulus of time is a 2-dimensional structure which describes the relevant (i.e. physical) degrees of freedom. In the 4D theory, the UV cut off is given by the string mass (evaluated near the big bang, i.e. near the BH). The UV string is wrapped around the black hole and it defines the minimum distance from the black hole singularity. However, what happens if we consider the same problem from the standpoint of chronon physics? The orbifold of time can be thought as a collection of chronons piled up together to form a string stretched between the branes. Since the radion is dual to the dilaton, we exploit the same picture with the radionic time and we have the collection of chronons (already mentioned above) wrapped around the singularity.  The chronons are not only the physical degrees of freedom of the theory, but they are also a (transplanckian) UV completion of the theory. The chronons are the fundamental {\it building blocks} of our model and, in particular, a 4D string is made of chronons.

Let us further elaborate these issues. When we move along the radionic string, at a certain angular coordinate, we reach the forbidden chronon and, if our intention is to believe in the causality principle locally, we are forced to change the Riemann sheet. In other words, we are forced to build a 3D structure. This seems to point out a F-theory origin of the model, but we do not know whether the 12th dimension is spacelike or timelike. Let us try with the simplest possibility: let us choose the 12th dimension to be spacelike. We pile up many annuli of time together and the direction orthogonal to the annuli is the 12th. The annuli are made of time and the free will of the observer (together with the chameleon mechanism) generates a large number of deviations in the timelike lines present on each annulus (for a recent discussion of free will and determinism, the reader is referred to \cite{Scardigli:libro}). Consequently, a Bravais lattice is forbidden in the annulus and, therefore, there is a macroscopic material which can be taken as a good representative of the time sheet: a standard glass (which is an amorphous solid).

We are building an architectural structure made of timelike lines. The vertical direction is chosen to be the 12th, it is spacelike and the definition of what we mean as ''vertical'' is related to the gravitational field of the black hole. As we will see, however, the vertical direction will be exchanged with the radial horizontal coordinate and, hence, the concept of ''vertical direction'' is more tricky than the one of our everyday life. Nevertheless, we obtain a 3-dimensional pillar where many annuli of glass are piled up together along the 12th coordinate. To proceed further, it is our intention to trust this analogy with the standard glass and, for this reason, we exploit some standard results of statistical physics. In particular, it is common knowledge that the glass is in equilibrium as a solid but only on time scales of roughly 10 years. Indeed, if we consider time intervals of roughly $10^3$ years, the glass behaves like a fluid (see e.g. the book \cite{Kardar:libro1, Kardar:libro2}). If we exploit this idea into our architectural structure, we infer that on very large time scales the chronons of the annuli of time behave like a fluid and they flake away from the annuli forming a sort of atmosphere  (i.e. the chronons stratify in harmony with the gravitational field). Consequently, for consistency reasons, it seems to be preferable a {\it timelike} 12-th dimension. Summarizing, we have a 3-dimensional architectural structure made of glassy time inserted in a sort of ''fog'' made of chronons. Remarkably, if we use a perfect gas approximation for the atmosphere of chronons, we obtain an {\it exponential} profile. We will further elaborate this point in the following paragraphs.

\subsubsection{Maxwellian analysis}

{\bf 2 dimensions}

We obtained a 3-dimensional scenario where timelike lines are organized in a geometrical glassy structure subjected to a fluidification process on very large time scales. Can we describe the distribution of chronons in this 3D time in a more quantitative way? We will proceed stepwise and we will start with the simpler 2D scenario. The starting geometry is the annulus already mentioned above. Now, let us cut the annulus along the direction of the forbidden chronon. In this way, we obtain a 2-dimensional figure homeomorphic to a rectangle ABCD. The edge AB corresponds to the UV radionic curve, the edges BC and AD correspond to the forbidden orbifold of time (i.e. to the orbifold of time whose angular coordinate is the forbidden chronon) and they should be identified together, the edge CD corresponds to the IR radionic curve. Now to the point. How can we obtain the distribution of chronons? One reasonable guess would be to exploit the Schroedinger equation. However, the model is not always quantum: as already mentioned above, the model is classical, for example, in the deep UV (at energies higher than the string mass near the big bang). A very physical way of connecting the UV with the IR is to exploit our relativity induced inflation. In relativity, two different observers read two different physical results in the same equation. Analogously, in our model, the UV observer and the IR observer interpret {\it the same equation} in two different ways: the IR observer exploits a quantum interpretation while the deep UV observer exploits a classical interpretation.

Now to the calculations. In the IR we expect the Schroedinger equation to be useful and the theory is free. Hence we write in 2D:
\begin{equation}
-\frac{1}{2} \nabla^2 \psi =0
\end{equation}
where $\psi$ is the wave function of the chronon in 2D and we have chosen $m=\hbar=1$. We emphasize that $\nabla^2=\partial_{t_1}^2+\partial_{t_2}^2$, hence it is not the standard Laplacian (we must exchange space and time). Remarkably the energy is set to zero: the UV-IR duality tells us that the same equation must be valid in the classical, non fluctuating, non-interacting regime, i.e. the deep UV.
In other words, the basic equation in the IR is quantum and it is formally given by 
\begin{equation}
\hat{H} \psi =0
\end{equation}
where $\hat{H}$ is the ''hamiltonian'' operator, while the basic equation in the deep UV is classical and it is given by
\begin{equation}
\nabla^2 \psi =0
\end{equation}
namely a Laplace equation in 2D.

The classical equation is discussed, for example, in classical electrodynamics (see \cite{Jackson:1998nia}, chapter 2). Now we choose the boundary conditions: $\psi=V$ at the edge AB (strong gravity at scales comparable to the string mass near the big bang), $\psi=0$ on the edges BC and AD (forbidden directions), $\psi=0$ on the edge CD (weak gravity). The conditions on the edges CD and AB are in harmony with the gradient of the gravitational field of the black hole along the orbifold of time. The solution is discussed in \cite{Jackson:1998nia} and it is
\begin{equation} 
\psi=\frac{2V}{\pi} Im[ln(\frac{1+Z}{1-Z})]
\label{2Ds}
\end{equation}
where we defined $Z=e^{i\pi/a}(t_R+i t_D)$ and $a$ is the length of AB, while $t_R$ and $t_D$ are respectively the radionic and dilatonic time. The logarithm is a multi-valued function in harmony with the interpretation of the floors of the Library as Riemann sheets. Remarkably, as mentioned in \cite{Jackson:1998nia}, in the IR the solution rapidly approaches
\begin{equation} 
\psi\longrightarrow \frac{4V}{\pi} e^{-\pi t_D/a} sin(\pi t_R/a).
\label{2Dsbis}
\end{equation}
The maxima and minima of the sine function correspond to the positions of the chronons along the radionic curve. The exponential factor is in harmony with our expectations for the dilaton from the MFM.
We mention once again that this solution must be interpreted classically in the deep UV but quantum mechanically in the IR. Basically we can start considering two contiguous chronons in the deep UV and we know that the system is classical. We solve the 2D Laplace equation and then we exploit the relativity induced inflation to map the solution of the equation into the quantum IR. Hence, a set of charts is obtained in the deep IR and in each chart the solution of \cite{Jackson:1998nia} is valid but it must be interpreted quantum mechanically as the wave function of the chronon. Two different observers read two different types of physics (classical and quantum) in the same equation. Needless to say, we explore regions beyond the UV cut-off of the SUGRA theory.

{\bf 3 dimensions}

Let us write the solution of the 3D Laplace equation in cylindrical coordinates. The cylinder corresponds to the basic pillar of the 3D glassy structure. Even if all the three dimensions are timelike, we will adopt a standard notation for the cylindrical coordinates, namely ($\rho, \phi, z$). The cylinder has a radius $a$ and a height $L$, the top and bottom surfaces being at $z=L$ and $z=0$. Once again we must choose the boundary conditions and we choose: $\psi=0$ on the side of the cylinder (in harmony with the boundary conditions in the IR of the annuli of time), $\psi=0$ at the bottom of the cylinder, $\psi=V(\rho, \phi)$ on the top of the cylinder.

If we assume that the solution is finite at $\rho=0$, we can write it in cylindrical coordinates as (see \cite{Jackson:1998nia})
\begin{equation} 
\psi(\rho, \phi, z)=\sum_{m=0}^\infty \sum_{n=1}^\infty J_m(k_{mn}\rho) sinh(k_{mn}z)(A_{mn} sin m \phi + B_{mn} cos m \phi),
\label{3Ds}
\end{equation}
where $k_{mn}=\frac{x_{mn}}{a}$ (m=1,2,3...) and $x_{mn}$ are the roots of $J_m(x_{mn})=0$.

The assumption of a finite solution in $\rho=0$ seems to clash with the presence of the bulk singularity. However, in the deep UV (namely at energies higher than the string mass near the big bang, namely at transplanckian energies) the curvature singularity becomes a coordinate singularity and, for this reason, the boundary condition is physical.

\subsubsection{Duality}

The purpose of this paragraph is to discuss the geometry of the 3D time.

Our theory is non-linear and, therefore, there are backreaction effects, namely effects from local physics to global one (in particular, effects from the fluctuating dilaton to the background one). However, these backreaction effects are important only when the harmonic approximation around a ground state is violated. To proceed further, let us consider various planetary systems or even various satellite systems. The average matter density and the typical length scale are obviously non-constant when we change the planetary system under discussion. In this way, we have various points (representative of matter densities of the various planetary/satellite systems) related together by an RG running (parametrizing the shift in the length scale). When we change the matter density, the chameleon field is shifted and the deviation from one ground state begins. When the non-linearity of the theory becomes important (i.e. when the deviation from the ground state is not small), S-duality applied to the RG running gives a non-trivial result: the ''old'' RG running towards the IR becomes a ''new'' RG running towards the UV. However, this new running will be S-dualized once again when we reach the ''boundary'' of the ground state, namely when the non-linearities are important (i.e. the harmonic approximation is not good anymore). Summarizing, we have a small oscillation of the dilaton around its ground state and this oscillation is interpreted as an RG running to the IR and then to the UV and then to the IR again (and so on...). We found an oscillating behaviour. An interesting line of development will compare this theoretical prediction with observational data of the Solar System.

One relevant question is: where is located the self-dual point of this S-duality in the effective lagrangian? It should map the radius of the Universe into the Planck mass: we choose a point ''in the middle'' and hence it should be located at a length scale roughly given by $10^{-5}$m. We define this length scale as {\it the lever of S-duality}. It is the self-dual length scale. We live on length scales comparable to one meter and our physics is obviously classical, however, we can S-dualize our world exploiting the lever and we obtain a bridge towards $10^{-10}$m where quantum mechanics is valid. S-duality maps classical physics into quantum physics and, in a sense, it replaces decoherence. However, this map classical-to-quantum is not the unique case: in the cosmological IR the model is quantum, but if we S-dualize the cosmological ground state we obtain the ground state in the UV at scales comparable to the string mass near the big bang. Later we will consider the geometry at transplanckian energies and, in that case, a conical geometry will be present while S-duality, once again, will map classical physics into quantum physics exchanging the tip of the cone with its base.

In order to describe the global geometry of the 3D time, let us exploit the comments just mentioned. We have a purely timelike glassy 3D structure (we have 3 times) and this structure is surrounded by a ''fog'' of chronons. As already mentioned above, if we use a perfect gas approximation for the chronons of the fog, we have an exponential profile. This is exactly the chameleonic profile characterizing the 4D lagrangian of the model. Remarkably, the glass becomes a fluid globally, but the exponential is in harmony with our local expectations (e.g. the local behaviour of the superpotential). Once again we find an example of global-to-local effects. To analyze the global geometry we use the lever of S-duality. In particular, in this way we can go out from the compactification space (it would have been better to say ''compactification time''). Let us further elaborate this issue. The radius of the fundamental glassy pillar is given by the distance between the two orbifold fixed points. However, S-duality produces a pillar whose radius is given by the distance between the black hole and the lever of S-duality. If we start from the black hole and we run to the IR, when we reach the lever at $10^{-5}$m
we enter into a dual space, i.e. we enter into another pillar and we exchange UV with IR. Now, we know that the dilaton and the radion are dual to each other, hence, we assume that the 3 times are dual to each other and, therefore, we can extend the glassy structure at infinity in the three timelike dimensions. We obtained the Borges' Library.

\begin{figure}
\centering
\includegraphics[width=1\textwidth]{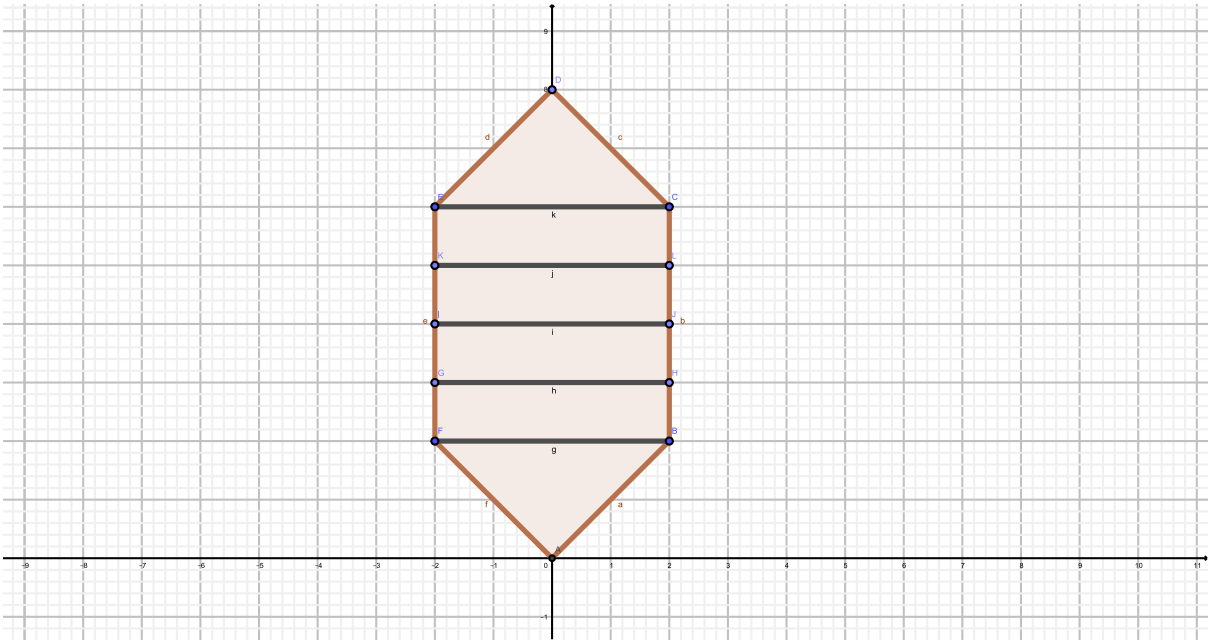}
\caption{Schematic representation of the basic cell of our 3-dimensional geometry. The vertical axis corresponds to the 12-th dimension. The horizontal segments $g,h,i,j,k$ represent the annuli of time. The dilaton has been dualized into the vertical direction. In particular, the UV (IR) cut off of the 4D theory corresponds to the base BF (CE) of the cone. In BF and in CE physics is quantum. In A physics is classical. The cone ABF corresponds to the deep UV comparable to the chronons' size. The cone CED is S-dual to the cone ABF. In A the theory is free, in BF (CE) the theory is strongly (weakly) coupled, in D the theory is strongly coupled. The 2-dimensional boundary is simply a 2-torus with one of the two cycles squeezed to a point in $A \equiv D$.}
\label{cilindro}
\end{figure}

Later we will search for some theoretical grounds supporting the duality among the 3 times. Now we simply assume that this duality is correct and, hence, we assume to put the orbifold of time along the vertical direction of the columns (this is not a problem because we can always go out from the compactification space exploiting the lever). Figure \ref{cilindro} shows a schematic view of our resulting geometrical configuration. S-duality, when applied between the tip (point A in fig. 1) and the base (segment BF in fig. 1) of the cone, becomes a map between classical and quantum physics. Remarkably, when we repeat the structure of figure \ref{cilindro} and we build the Library, we can also cut the geometry in the midpoint of the segment FE (or BC): in this case the resulting fundamental cell of geometry has a hourglass structure. The Library becomes reminiscent of those games for children where the sand (i.e. the chronons flaked away from the glassy time in our model) go through a certain number of hourglasses and passages.

\begin{figure}
\centering
\includegraphics[width=1\textwidth]{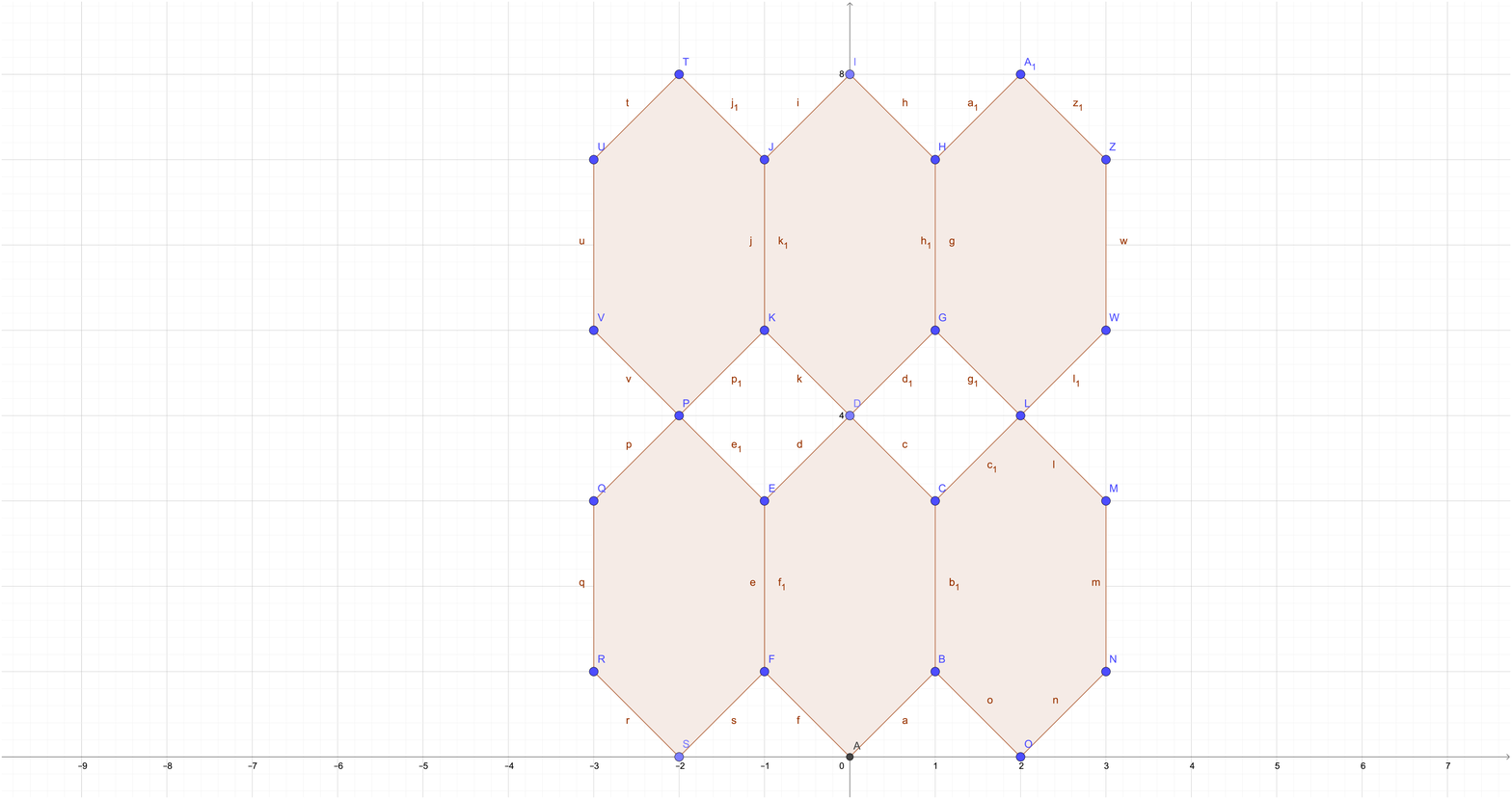}
\caption{A schematic view of the Library.}
\label{library}
\end{figure}

Let us add some remarks.
\begin{itemize}
\item  A type IIB theory on orbifold with complex dilaton equal to the complex structure of the annulus corresponds to F-theory on the Library (i.e. on the fibre bundle), see e.g. \cite{Sen:1996vd}. If we use the vertical orbifold parity we obtain the worldsheet parity of type IIB string, namely we have type I open strings. One might think that it is a nonsense to put the type IIB string along a timelike direction, however, when we quantize gravity there is no difference between spacelike and timelike intervals. Evolving this system in time, we obtain a global string. Once again the global string is made of chronons and the worldsheet parity of the global vertical string corresponds to the parity of the vertical orbifold.
\item Remarkably, there are cavity walls among the columns. It is not possible to go through the cavity wall, but the Library can rotate depending on our choice of radionic (angular) coordinate: the choice of radionic coordinate fixes the direction where it is possible to leave the compactification space. Figure \ref{library} shows a sample of the Library (naturally the pattern should be extended to infinity). The hourglass-like structure is manifest.
\item The glass is amorphous (''disordered'' in a sense) but when we repeat the same glassy structure we obtain a geometrical order (analogous to the Order of Borges). However this ordered structure will become a fluid and the instant of time that we are living right now might become, in due time, part of the atmosphere of the Library and we already perceive the presence of this atmosphere of chronons in the exponential function of the 4D theory which is characterizing the masses of the chameleon fields (i.e. of the chronons in their foggy atmosphere). Once again global-to-local effects are important.
\end{itemize}
Interestingly, the Library is obtained from S-duality: duality defines geometry while geometry produces global structures and these global structures tell us something about local physics.

\section{Phenomenology}
\subsection{Quantum formation of planets and satellites}
\label{Qplanets}

As already mentioned above, in our M-theory model quantum mechanics is stretched to very large distances. This creates interesting connections between the chronon and planetary physics. Indeed, the cosmological scale factor is an exponential function of the dilaton and this plays the role of an exponential ``magnifying lens'' for the quantum fluctuations. When the matter density is small, the dilaton leads to an exponential enhancement of the fluctuations.

\subsubsection{Matching the equations}

Now it is time to link our modified Schroedinger equation with the equation of \cite{Scardigli:2005fr}.

We start from \ref{mod}, but we write it reinstalling the fundamental constants. Hence we write our equation for a wave function $\Psi$ as
\bea
i \hbar \partial_t \Psi= -\frac{\hbar^2 \nabla^2}{2m} \Psi +\frac{\lambda}{12} \hbar^3 \frac{e^{-2imc^2 t/\hbar}}{m^2 c} \Psi^3.
\eea
The next step is to linearize this equation assuming a real wave function. We write
\be
\Psi^3 = \Psi^2 \Psi = \rho_p \Psi
\ee
where $\rho_p$ is the probability density for a single matter particle. What should we write for the probability density? This is a non-trivial problem, because in order to write the wave function we must solve the equation. We suggest to interpret the $\rho_p$ term as
\be
\rho_p=\Psi_{pilot}^2
\ee
where $\Psi_{pilot}$ is a sort of bohmian pilot wave which plays the role of a potential for particles. Since the central object is approximately spherical and the only interaction we are considering is gravity, we assume a gravitational behaviour of the pilot wave, namely we assume that (forgetting constant prefactors of properly chosen units)
\be
\Psi_{pilot}\simeq 1/r. 
\ee
 The normalization condition can be written as
\be
1= \int_V d^3 x \rho_p \simeq \int_v d^3 x \rho_p 
\ee
where $V$ is the volume of the universe and $v$ is the volume of the solar system (or the volume of the satellite system we are interested in). In other words, if we are interested in planets, we normalize the wave function on a sphere of radius $d$ comparable to the size of the system. Hence we write
\be
\rho_p =\chi \frac{1}{r^2 d}
\ee
where $\chi$ is a dimensionless constant.
Therefore we rewrite the last term of our equation in a linearized form as
\be
\frac{\lambda}{12} \frac{\hbar^3}{m^2 c} e^{-2 i m c^2 t/\hbar} \frac{1}{r^2 d} \Psi
\ee
where $\chi$ has been absorbed into $\lambda$.
This term can easily produce the contribution of the operator ${\cal {\hat P}}$ of \cite{Scardigli:2005fr}, indeed, if we assume a quantization of time then we can write 
\be t= p t_{ch}
\ee
where $p$ is an integer counting the number of chronons, while $t_{ch}$ is the fundamental unit of time (the chronon). Obviously we have $p>>1$.
The link with \cite{Scardigli:2005fr} is obtained exploiting the orbifold of time. Indeed, let us apply the $\mathbb{Z}_2$ parity of the orbifold on the linearized version of equation \ref{mod} to obtain
\bea
-i \hbar \partial_t \Psi= -\frac{\hbar^2 \nabla^2}{2m} \Psi + \frac{\lambda}{12} \frac{\hbar^3}{m^2 c} e^{2 i m c^2 t/\hbar} \frac{1}{r^2 d} \Psi.
\eea
To proceed further, we work with Wick-rotated time $t_E$ (in harmony with the bag model of reference \cite{Zanzi:2006xr}) and, therefore, we rewrite the oscillating exponential phase as a real exponential, namely
\bea
 \hbar \partial_{t_E} \Psi= -\frac{\hbar^2 \nabla^2}{2m} \Psi +\frac{\lambda}{12} \frac{\hbar^3}{m^2 c} e^{2  m c^2 t_E/\hbar} \frac{1}{r^2 d} \Psi.
\eea
Now we change notations in harmony with reference \cite{Scardigli:2005fr}, namely we define $s=\hbar/m$ and we write the hamiltonian {\it per unit mass} as
\be
H_M=-\frac{s^2 \nabla^2}{2}  +i^2\frac{(- \lambda )}{12} \frac{s^3 }{ c} e^{2  m c^2 p (t_{ch})_E/\hbar} \frac{1}{r^2 d}. 
\ee
The relevant term is the last one, we rewrite it as
\be
- \frac{s^2}{2r^2}\frac{(-\lambda )}{6} \frac{s }{ c} e^{2  m c^2 p (t_{ch})_E/\hbar} \frac{1}{ d} 
\ee
and, consequently, the operator ${\cal {\hat P}}_\phi^2$ of \cite{Scardigli:2005fr} is linked to our model by the following dictionary\\

{\bf Dictionary:}\\
\bea
{\cal{\hat P}}^2_\phi=[i e^{-i \lambda_S \partial_\phi}]^2 \Longleftrightarrow \frac{(-\lambda )}{6} \frac{s }{ cd} e^{2  m c^2 p ({\hat t}_{ch})_E/\hbar} ;\\
 \lambda_S \Longleftrightarrow   mc^2 (t_{ch})_E/\hbar. 
\eea

$\Box$\\

In this way the action of the operator on the angular part of the wave function $e^{ip \phi}$ provides the same eigenvalue of the work \cite{Scardigli:2005fr}, indeed the formula of \cite{Scardigli:2005fr} gives
\be
{\cal{\hat P}}^2_\phi e^{ip \phi}=[i e^{p \lambda_S}]^2 e^{ip \phi}=-e^{2p \lambda_S} e^{ip \phi}
\ee
while our dictionary tells us to write the eigenvalue equation
\be
\frac{(-\lambda )}{6} \frac{s }{ cd} e^{2  m c^2 p ({\hat t}_{ch})_E/\hbar} e^{ip \phi}= \frac{(-\lambda )}{6} \frac{s }{ cd} e^{2  m c^2 p ( t_{ch})_E/\hbar} e^{ip \phi}.
\ee
As we see the link imposes a peculiar choice on $\lambda$, namely
\be
\frac{(-\lambda )}{6} \frac{s }{ cd}=-1
\ee
or, equivalently, in units $\hbar=c=1$
\be
\lambda=6md.
\ee

  In \cite{Scardigli:2005fr}  the standard term of the Schroedinger equation
\be
$S$=-\frac{\hbar^2}{2r^2} \frac{\partial^2}{\partial \phi^2}
\ee
is replaced with the new one related to the operator ${\cal {\hat P}}$.

We can check explicitly that, in the MFM, the standard term $S$ is subleading with respect to the new one, because the required suppression is due to the exponential factor with Wick-rotated time and, hence, we recover the equation of \cite{Scardigli:2005fr}.

\subsubsection{Titius-Bode law and Io: one possible interpretation}

As we mentioned above, the number of chronons $p$ must be identified with the quantum number of reference \cite{Scardigli:2005fr}. Naturally the number of chronons is extremely large and, therefore, our M-theory equation must be exploited in the limit of large $p$. Remarkably, the compactification of time on the $S^1/\mathbb{Z}_2$ orbifold tells us that the sign of (the dilatonic) time is irrelevant. We infer that the limit of large $p$ is simply the limit $p\longrightarrow \pm \infty$. This is precisely the limit considered in reference \cite{Scardigli:2005fr} to discuss, in the case of plus sign, the Titius-Bode law and, in the case of minus sign, the orbit of the innermost satellites. In particular, as already discussed in \cite{Scardigli:2005fr}, the agreement with the orbital radius of Io (the innermost satellite of Jupiter) is particularly good.

Here is one of the possible interpretations of these results. Today we can use classical physics in the solar system because the exponential magnifying lens is so effective that the wavelength of quantum fluctuations is much larger than 10 AU. In this case the fluctuation is not observable and classical physics is a good approximation. However, before the formation of the solar system, namely a few billions of years ago, the universe was smaller, the cosmological dilaton was heavier and, hence, the exponential magnifying lens was less effective. Consequently, we can imagine a {\it quantum} scenario for the formation of planets and satellites characterized by quantum fluctuations stretched to length scales comparable to the size of the ''primordial'' Solar System. Following this idea, let us start considering a (roughly) spherical distribution of matter that will evolve into our Sun. Let us assume that this spherical distribution of matter does not host nuclear reactions. In other words, the Sun is not yet switched on. We will also assume that this Sun is rotating and, hence, the rotational symmetry is broken but an axial symmetry is still present. Naturally, this Sun will attract gravitationally  the matter particles: we have a gravitational collapse of matter towards the Sun. Let us split the problem into a collection of beams of matter particles collapsing radially towards the Sun.
Now to the point. Classically we know that matter particles will fall into the Sun. Our intention is to find a mechanism to generate circular rings of matter around the Sun (or around Saturn, Jupiter...) and then to interpret these rings as the regions where planets (or satellites) will be present. The main idea is to modify the radial collapse that we expect classically. What happens quantum mechanically? In the MFM a beam of matter particles falling towards the Sun is analogous to a beam of electrons traveling towards a circular hole. Indeed, when the Sun does not yet host nuclear reactions, the background photons are screened by the Sun and, hence, a spherical shadow is projected on the beam. In this shadow the energy density of photons is smaller and, therefore, the chameleon mechanism (see also \cite{Zanzi:2012ha, Zanzi:2015cch}) tells us that the geometry of the problem is completely analogous to a plane wave traveling towards a circular hole. As we know circular rings will be formed whenever the quantum interference is constructive. Here we expect a similar phenomenon: circular rings of matter will be formed in the equatorial plane of the Sun but only where quantum interference is constructive. Remarkably, the formation of rings in the equatorial plane is related to the symmetry properties of the ground state. Indeed, the Sun is rotating and, therefore, it is not perfectly spherical. This symmetry property of the matter distribution is transferred to the ground state exploiting the chameleon mechanism and, hence, the rotation of the Sun selects the equatorial plane as the region of constructive quantum interference. Had we neglected the rotation of the Sun, we would have formed $S^2$ branes of matter rotating around the Sun. This point should be further elaborated along the lines of \cite{Zanzi:2015evj, Zanzi:2016yrg}. When matter particles are falling towards the Sun, the role of the screen in played by the gravitational field of the Sun. This gravitational field shares the same symmetry properties of the Sun: we have an axial symmetry. The chameleon mechanism feels the presence of the gravitational field because gravity is summarized by a conformal anomaly in harmony with the CEP of reference \cite{Zanzi:2015evj}. This anomaly induces a chameleonic jump from the initial ground state to a final one and transfers the symmetry of the Sun (which is not spherical but only axial) to the final ground state. Hence, rotational symmetry is spontaneously broken and a quantum deviation is obtained, but an axial symmetry is still present and the equatorial plane is selected.

Some useful related references are \cite{Jizba:2014taa, Blasone:2009ky, Blasone:2009yp}.

\subsection{Oscillating data}

 Let us start considering once again the Maxwellian analysis in 2D and in 3D. These two analysis should be compatible with each other: the solution of the 3D Laplace equation on the top and bottom of the cylinder should match the 2D solution. Let us see what we can infer from this compatibility. The 2D solution mentioned above

\begin{equation} 
\psi=\frac{2V}{\pi} Im[ln(\frac{1+Z}{1-Z})]
\label{2Ds}
\end{equation}
 while the 3D one is
\begin{equation} 
\psi(\rho, \phi, z)=\sum_{m=0}^\infty \sum_{n=1}^\infty J_m(k_{mn}\rho) sinh(k_{mn}z)(A_{mn} sin m \phi + B_{mn} cos m \phi).
\label{3Ds}
\end{equation}

The compatibility of the two solutions supports the identification of an exponential behaviour (which characterizes the dilatonic coordinate in the 2D solution) with the oscillating behavior of the Bessel functions (which characterizes the radial coordinate of the cylinder in the 3D solution). This is in harmony with the S-dualized RG running mentioned above. Indeed, the exponential potential of the 4D lagrangian is related to the matter density by the chameleon mechanism, because the matter density selects the ground state of the dilaton through the conformal anomaly. Consequently, if we shift the matter density we have a RG running of the dilaton summarized by a motion along the exponential potential. If we S-dualize this RG running, we are led to an oscillating behaviour. Consequently, S-duality is the theoretical ingredient which makes this compatibility easier to understand. 

Remarkably, {\it the duality among the three times is supported by the Maxwellian analysis}, because the profile of the 3D solution along the longitudinal direction of the cylinder is exponential like the profile of the 2D solution along the orbifold direction of the annulus.

Let us conclude with a more phenomenological comment. Let us consider the GW signal of VIRGO \cite{Abbott:2016blz}. We have the inspiral and merger of two black holes. Hence, these two objects become closer and closer to each other, the system explores decreasing length scales and, therefore, increasing matter densities. In our model this process is described by an RG running to the UV region. Exploiting S-duality, the exponential shift of the ground state is mapped into an oscillating behaviour around the ground state. When the GW is detected by VIRGO, we imagine a quantization of gravity and (neglecting the gravitational field of the Earth, which is not relevant for our discussion) we are left with the gravitational field described by an oscillation of the dilaton around the ground state. The dilaton propagates from the source to the detector and it brings to VIRGO the memory of the oscillation present at the source (i.e. the RG running at the source). In the ground state where we quantize gravity (namely the ground state of VIRGO), the Lorentz group is not physically relevant anymore and, therefore, the oscillation produces oscillating violations of Local Lorentz Invariance. Therefore, in this model, we understand the strain as ($\delta L_x - \delta L_y$), namely as a phenomenon due to the oscillating breakdown of Local Lorentz Invariance at the location of the VIRGO experiment.

In our model, the oscillation detected by VIRGO is a direct consequence of S-duality and it should be described in terms of Bessel functions.

However, things are not so easy. Indeed, we know that a GW is described by classical GR and the spin related to the metric is 2. How can we map this spin-2 information of the gravitational wave into a scalar dilaton? The best answer at this stage is that in this model the concept of spin is not relevant anymore when we quantize gravity on the brane. As already pointed out above, these surprising results concerning spin are extracted from the lagrangian. However, we are well aware of the fact that it might be that our lagrangian has nothing to do with the real world.

\section{Conclusions}

Some aspects of chronon physics have been discussed from the standpoint of M-theory and F-theory exploiting a model originally proposed in \cite{Bergshoeff:2000zn}. Remarkably, the causality principle has been violated globally and, for this reason, we are led to a deeper question: if we get rid of the causality principle, what is the more fundamental principle we trust? At this stage, probably, the answer is Geometry. Geometry seems to be a better guideline than causality (at least globally) and, in this sense, our paper should be considered conceptually as a mathematics, not physics, paper.

Ghosts and unitarity violations are typically a problematic aspect of 2 time physics. In our case, however, the second and the third time are fixed at our energy scales, hence these problems are solved as far as all practical purposes are concerned. In our model we do not have CTC and hence there is no problem in considering the $S^4$ manifold with timelike coordinates included. When we quantize gravity spacelike and timelike intervals are equivalent to each other because QG effects can shift the position of the horizons of the black hole. We are left with a very familiar 4D theory with euclidean time that we take as our starting point for future research.

We found a surprising spin-independence in this model. The concept of spin seems not relevant anymore when we quantize gravity on the brane. Our strategy here is to think that physics is contained in the lagrangian. If we try to make things always intuitive, it might be a mistake. These surprising results concerning spin are extracted from the lagrangian. However, we are well aware of the fact that it might be that our lagrangian has nothing to do with the real world.

Our scenario has non-trivial overlaps with some papers of the literature. The quantum of time has been discussed, for example, in \cite{Caldirola:1977fu}. Many authors tried to use a Schroedinger-like equation to derive the Titius-Bode law (e.g. see \cite{Caswell:1929aa, Albeverio:1983aa}). A promising line of development will investigate potential connections with \cite{Wilczek:2012jt, Caswell:1929aa, Albeverio:1983aa, Freidel:2003sp, Nelson:1985aa, Godel:1949ga, Gama:2017eip, Amendola:2007qs, Sornette:1997pb, Barbour:2011dn}. 2-time physics has been discussed in \cite{Araya:2013bca, Bars:2010zz} and references therein. Other useful references are \cite{Anderson:2017jij, Mersini-Houghton:2012mha} and \cite{Jizba:2014taa, Blasone:2009ky, Blasone:2009yp}.

\subsection*{Acknowledgements}
Special thanks are due to Ignatios Antoniadis, Daniele Bertacca, Sergio Luigi Cacciatori, Gino Isidori, Claus Kiefer, Massimo Pietroni, Fabio Scardigli for useful conversations. Special thanks are also due to Mauro Moretti for his help with (and his Lecture Notes on) coherent states of Quantum Mechanics.


\providecommand{\href}[2]{#2}\begingroup\raggedright\endgroup


\begin{thebibliography}{10}

\bibitem{Straumann:2013ns}
N.~Straumann, {\em {General Relativity - 2nd edition}}.
\newblock
2013.
\newblock

\bibitem{Hamber:2009zz}
H.~W. Hamber, ``{Quantum gravitation: The Feynman path integral approach},''
{\em Springer-Verlag} (2009)  1--342.

\bibitem{Kiefer:2007ria}
C.~Kiefer, {\em {Quantum Gravity}}.
\newblock Oxford University Press, New York,
2007.
\newblock

\bibitem{Rovelli:2004tv}
C.~Rovelli, {\em {Quantum gravity}}.
\newblock Cambridge Monographs on Mathematical Physics. Univ. Pr., Cambridge,
  UK, 2004.

\bibitem{Blumenhagen:2013fgp}
R.~Blumenhagen, D.~Luest, and S.~Theisen, ``{Basic concepts of string
  theory},''
{\em Springer-Verlag} (2013)  1--782.

\bibitem{Becker:2007zj}
K.~Becker, M.~Becker, and J.~H. Schwarz, ``{String theory and M-theory: A
  modern introduction},''. Cambridge, UK: Cambridge Univ. Pr. (2007) 739 p.

\bibitem{Ovrut:2002hi}
B.~A. Ovrut, ``{Lectures on heterotic M theory},''
\href{http://arxiv.org/abs/hep-th/0201032}{{\tt arXiv:hep-th/0201032
  [hep-th]}}.

\bibitem{Krause:1900zz}
A.~Krause, ``{Heterotic M-theory and cosmology},'' in {\em {In *Erdmenger, J.
  (ed.): String cosmology* 267-303}}.
\newblock
2009.
\newblock

\bibitem{Waldram:2001my}
D.~Waldram, ``{Introduction to model building in heterotic M-theory},''
  pp.~763--850.
\newblock
2001.
\newblock

\bibitem{Zanzi:2016thx}
A.~Zanzi, ``{The cosmological constant problem in heterotic-M-theory and the
  orbifold of time},''
\href{http://arxiv.org/abs/1607.04195}{{\tt arXiv:1607.04195 [hep-th]}}.

\bibitem{Fujii:2002sb}
Y.~Fujii, ``{Mass of the dilaton and the cosmological constant},''
  \href{http://dx.doi.org/10.1143/PTP.110.433}{{\em Prog. Theor. Phys.} {\bf
  110} (2003)  433--439},
\href{http://arxiv.org/abs/gr-qc/0212030}{{\tt arXiv:gr-qc/0212030}}.

\bibitem{Bergshoeff:2000zn}
E.~Bergshoeff, R.~Kallosh, and A.~Van~Proeyen, ``{Supersymmetry in singular
  spaces},'' \href{http://dx.doi.org/10.1088/1126-6708/2000/10/033}{{\em JHEP}
  {\bf 10} (2000)  033},
\href{http://arxiv.org/abs/hep-th/0007044}{{\tt arXiv:hep-th/0007044
  [hep-th]}}.

\bibitem{Zanzi:2010rs}
A.~Zanzi, ``{Chameleonic dilaton, nonequivalent frames, and the cosmological
  constant problem in quantum string theory},''
  \href{http://dx.doi.org/10.1103/PhysRevD.82.044006}{{\em Phys. Rev.} {\bf
  D82} (2010)  044006},
\href{http://arxiv.org/abs/1008.0103}{{\tt arXiv:1008.0103 [hep-th]}}.

\bibitem{Zanzi:2015evj}
A.~Zanzi, ``{Chameleonic equivalence postulate and wave function collapse},''
  {\em Electron. J. Theor. Phys.} {\bf 12} (2015) no.~33, 1--28,
\href{http://arxiv.org/abs/1404.1942}{{\tt arXiv:1404.1942 [quant-ph]}}.

\bibitem{Zanzi:2014aia}
A.~Zanzi and B.~Ricci, ``{Chameleon fields and solar physics},''
  \href{http://dx.doi.org/10.1142/S0217732315500534,
  10.1142/S0217732316920012}{{\em Mod. Phys. Lett.} {\bf A30} (2015) no.~10,
  1550053}, \href{http://arxiv.org/abs/1405.1581}{{\tt arXiv:1405.1581
  [hep-ph]}}.
[Erratum: Mod. Phys. Lett.A31,no.4,1692001(2016)].

\bibitem{Zanzi:2016yrg}
A.~Zanzi, ``{Quantum mechanics before the big bang in heterotic-M-theory},''
  {\em Electron. J. Theor. Phys.} {\bf 13} (2016) no.~36, 13--24,
\href{http://arxiv.org/abs/1607.04369}{{\tt arXiv:1607.04369 [hep-th]}}.

\bibitem{Zanzi:2012ha}
A.~Zanzi, ``{Species, chameleonic strings and the concept of particle},''
\href{http://arxiv.org/abs/1206.4794}{{\tt arXiv:1206.4794 [hep-th]}}.

\bibitem{Zanzi:2015cch}
A.~Zanzi, ``{Chameleonic Theories: A Short Review},''
  \href{http://dx.doi.org/10.3390/universe1030446}{{\em Universe} {\bf 1}
  (2015)  446--475},
\href{http://arxiv.org/abs/1602.03869}{{\tt arXiv:1602.03869 [gr-qc]}}.

\bibitem{Khoury:2003aq}
J.~Khoury and A.~Weltman, ``{Chameleon fields: Awaiting surprises for tests of
  gravity in space},''
  \href{http://dx.doi.org/10.1103/PhysRevLett.93.171104}{{\em Phys. Rev. Lett.}
  {\bf 93} (2004)  171104},
\href{http://arxiv.org/abs/astro-ph/0309300}{{\tt arXiv:astro-ph/0309300}}.

\bibitem{Khoury:2003rn}
J.~Khoury and A.~Weltman, ``{Chameleon cosmology},''
  \href{http://dx.doi.org/10.1103/PhysRevD.69.044026}{{\em Phys. Rev.} {\bf
  D69} (2004)  044026},
\href{http://arxiv.org/abs/astro-ph/0309411}{{\tt arXiv:astro-ph/0309411}}.

\bibitem{Brax:2004ym}
P.~Brax, C.~van~de Bruck, and A.~C. Davis, ``{Is the radion a chameleon?},''
  \href{http://dx.doi.org/10.1088/1475-7516/2004/11/004}{{\em JCAP} {\bf 0411}
  (2004)  004},
\href{http://arxiv.org/abs/astro-ph/0408464}{{\tt arXiv:astro-ph/0408464}}.

\bibitem{Zanzi:2012du}
A.~Zanzi, ``{Chameleonic dilaton and conformal transformations},''
\href{http://arxiv.org/abs/1206.4463}{{\tt arXiv:1206.4463 [hep-th]}}.

\bibitem{Zanzi:2006xr}
A.~Zanzi, ``{Neutrino dark energy and moduli stabilization in a BPS braneworld
  scenario},'' \href{http://dx.doi.org/10.1103/PhysRevD.73.124010}{{\em Phys.
  Rev.} {\bf D73} (2006)  124010},
\href{http://arxiv.org/abs/hep-ph/0603026}{{\tt arXiv:hep-ph/0603026}}.

\bibitem{Zanzi:2012bf}
A.~Zanzi, ``{Dilaton stabilization and composite dark matter in the string
  frame of heterotic-M-theory},''
\href{http://arxiv.org/abs/1210.4615}{{\tt arXiv:1210.4615 [hep-th]}}.

\bibitem{Horava:1995qa}
P.~Horava and E.~Witten, ``{Heterotic and type I string dynamics from eleven
  dimensions},'' \href{http://dx.doi.org/10.1016/0550-3213(95)00621-4}{{\em
  Nucl. Phys.} {\bf B460} (1996)  506--524},
\href{http://arxiv.org/abs/hep-th/9510209}{{\tt arXiv:hep-th/9510209}}.

\bibitem{Horava:1996ma}
P.~Horava and E.~Witten, ``{Eleven-Dimensional Supergravity on a Manifold with
  Boundary},'' \href{http://dx.doi.org/10.1016/0550-3213(96)00308-2}{{\em Nucl.
  Phys.} {\bf B475} (1996)  94--114},
\href{http://arxiv.org/abs/hep-th/9603142}{{\tt arXiv:hep-th/9603142}}.

\bibitem{Dorigoni:2009ra}
D.~Dorigoni and V.~S. Rychkov, ``{Scale Invariance + Unitarity $=>$ Conformal
  Invariance?},''
\href{http://arxiv.org/abs/0910.1087}{{\tt arXiv:0910.1087 [hep-th]}}.

\bibitem{Araya:2013bca}
I.~J. Araya and I.~Bars, ``{Generalized dualities in one-time physics as
  holographic predictions from two-time physics},''
  \href{http://dx.doi.org/10.1103/PhysRevD.89.066011}{{\em Phys. Rev.} {\bf
  D89} (2014) no.~6, 066011},
\href{http://arxiv.org/abs/1311.4205}{{\tt arXiv:1311.4205 [hep-th]}}.

\bibitem{Bars:1997bz} 
  I.~Bars and C.~Kounnas,
  ``Theories with two times,''
  Phys.\ Lett.\ B {\bf 402}, 25 (1997)
  doi:10.1016/S0370-2693(97)00452-8
  [hep-th/9703060].

\bibitem{Bars:2006dy} 
  I.~Bars,
  ``The Standard Model of Particles and Forces in the Framework of 2T-physics,''
  Phys.\ Rev.\ D {\bf 74}, 085019 (2006)
  doi:10.1103/PhysRevD.74.085019
  [hep-th/0606045].

\bibitem{Bars:2010zz}
I.~Bars and J.~Terning, ``{Extra dimensions in space and time},''
{\em New York, USA: Springer (2010) 217 p} (2010)  .

\bibitem{Gasperini:libro}
M.~Gasperini, ``{Elements of string cosmology},''. Cambridge, Univ. Pr. (2007)
  552 p.

\bibitem{Cohen:1998zx}
A.~G. Cohen, D.~B. Kaplan, and A.~E. Nelson, ``{Effective field theory, black
  holes, and the cosmological constant},''
  \href{http://dx.doi.org/10.1103/PhysRevLett.82.4971}{{\em Phys. Rev. Lett.}
  {\bf 82} (1999)  4971--4974},
\href{http://arxiv.org/abs/hep-th/9803132}{{\tt arXiv:hep-th/9803132
  [hep-th]}}.

\bibitem{Wang:2016och}
S.~Wang, Y.~Wang, and M.~Li, ``{Holographic Dark Energy},''
  \href{http://dx.doi.org/10.1016/j.physrep.2017.06.003}{{\em Phys. Rept.} {\bf
  696} (2017)  1--57},
\href{http://arxiv.org/abs/1612.00345}{{\tt arXiv:1612.00345 [astro-ph.CO]}}.

\bibitem{Preparata:1997ap}
G.~Preparata, S.~Rovelli, and S.~S. Xue, ``{Gas of wormholes: A Possible ground
  state of quantum gravity},''
  \href{http://dx.doi.org/10.1023/A:1001992900070}{{\em Gen. Rel. Grav.} {\bf
  32} (2000)  1859--1931},
\href{http://arxiv.org/abs/gr-qc/9806044}{{\tt arXiv:gr-qc/9806044 [gr-qc]}}.

\bibitem{Fujii:2003pa}
Y.~Fujii and K.~Maeda, ``{The scalar-tensor theory of gravitation},''.
  Cambridge Univ. Press (2003) 240 p.

\bibitem{Brax:2002nt}
P.~Brax, C.~van~de Bruck, A.~C. Davis, and C.~S. Rhodes, ``{Cosmological
  evolution of brane world moduli},''
  \href{http://dx.doi.org/10.1103/PhysRevD.67.023512}{{\em Phys. Rev.} {\bf
  D67} (2003)  023512},
\href{http://arxiv.org/abs/hep-th/0209158}{{\tt arXiv:hep-th/0209158}}.

\bibitem{Baumann:2014nda}
D.~Baumann and L.~McAllister, {\em {Inflation and String Theory}}.
\newblock Cambridge University Press, 2015.
\newblock \href{http://arxiv.org/abs/1404.2601}{{\tt arXiv:1404.2601
  [hep-th]}}.

\bibitem{Cohen:2005cra}
C.~Cohen-Tannoudji, B.~Diu, and F.~Laloe, ``{Quantum mechanics}'', (1977) Hermann - Paris.

\bibitem{Blumenhagen:2009zz}
R.~Blumenhagen and E.~Plauschinn, ``{Introduction to conformal field theory},''
  {\em Lect. Notes Phys.} {\bf 779} (2009)  1--256.

\bibitem{Nieto:libro}
M.~M. Nieto, ``{The Titius-Bode law of planetary distances - its history and
  theory},''. Pergamon Press (1972) 161 p.

\bibitem{Scardigli:2005fr}
F.~Scardigli, ``{A quantum-like description of the planetary systems},''
  \href{http://dx.doi.org/10.1007/s10701-007-9151-7}{{\em Found. Phys.} {\bf
  37} (2007)  1278--1295},
\href{http://arxiv.org/abs/gr-qc/0507046}{{\tt arXiv:gr-qc/0507046 [gr-qc]}}.

\bibitem{Nikolic:2012wj}
H.~Nikolic, ``{Relativistic Quantum Mechanics and Quantum Field Theory},''
\href{http://arxiv.org/abs/1205.1992}{{\tt arXiv:1205.1992 [hep-th]}}.

\bibitem{Gutperle:2002ai}
M.~Gutperle and A.~Strominger, ``{Space - like branes},''
  \href{http://dx.doi.org/10.1088/1126-6708/2002/04/018}{{\em JHEP} {\bf 04}
  (2002)  018},
\href{http://arxiv.org/abs/hep-th/0202210}{{\tt arXiv:hep-th/0202210
  [hep-th]}}.

\bibitem{Gourgoulhon:2013qva}
E.~Gourgoulhon, {\em {Special relativity in general frames}}.
\newblock Springer, 2013.

\bibitem{Caldirola:1977fu}
P.~Caldirola, ``{Chronon in Quantum Theory},''
\href{http://dx.doi.org/10.1007/BF02785060}{{\em Lett. Nuovo Cim.} {\bf 18}
  (1977)  465--468}.

\bibitem{Weinberg:libro}
S.~Weinberg, ``{The quantum theory of fields (Volume 1)},''. Cambridge
  University Press (1995) 609 p.

\bibitem{Kachru:2003aw}
S.~Kachru, R.~Kallosh, A.~D. Linde, and S.~P. Trivedi, ``{De Sitter vacua in
  string theory},'' \href{http://dx.doi.org/10.1103/PhysRevD.68.046005}{{\em
  Phys.Rev.} {\bf D68} (2003)  046005},
\href{http://arxiv.org/abs/hep-th/0301240}{{\tt arXiv:hep-th/0301240
  [hep-th]}}.

\bibitem{Cheung:2007st}
C.~Cheung, P.~Creminelli, A.~L. Fitzpatrick, J.~Kaplan, and L.~Senatore, ``{The
  Effective Field Theory of Inflation},''
  \href{http://dx.doi.org/10.1088/1126-6708/2008/03/014}{{\em JHEP} {\bf 03}
  (2008)  014},
\href{http://arxiv.org/abs/0709.0293}{{\tt arXiv:0709.0293 [hep-th]}}.

\bibitem{Ortin:2015hya}
T.~Ortin, {\em {Gravity and Strings - 2nd edition}}.
\newblock Cambridge University Press, 2015.

\bibitem{Witten:2012bh}
E.~Witten, ``{Superstring Perturbation Theory Revisited},''
\href{http://arxiv.org/abs/1209.5461}{{\tt arXiv:1209.5461 [hep-th]}}.

\bibitem{Hori:2003ic}
K.~Hori, S.~Katz, A.~Klemm, R.~Pandharipande, R.~Thomas, C.~Vafa, R.~Vakil, and
  E.~Zaslow, {\em {Mirror symmetry}}, vol.~1 of {\em Clay mathematics
  monographs}.
\newblock AMS, Providence, USA,
2003.
\newblock

\bibitem{Vafa:1996xn}
C.~Vafa, ``{Evidence for F theory},''
  \href{http://dx.doi.org/10.1016/0550-3213(96)00172-1}{{\em Nucl. Phys.} {\bf
  B469} (1996)  403--418},
\href{http://arxiv.org/abs/hep-th/9602022}{{\tt arXiv:hep-th/9602022
  [hep-th]}}.

\bibitem{Morrison:1996na}
D.~R. Morrison and C.~Vafa, ``{Compactifications of F theory on Calabi-Yau
  threefolds. 1},'' \href{http://dx.doi.org/10.1016/0550-3213(96)00242-8}{{\em
  Nucl. Phys.} {\bf B473} (1996)  74--92},
\href{http://arxiv.org/abs/hep-th/9602114}{{\tt arXiv:hep-th/9602114
  [hep-th]}}.

\bibitem{Morrison:1996pp}
D.~R. Morrison and C.~Vafa, ``{Compactifications of F theory on Calabi-Yau
  threefolds. 2.},'' \href{http://dx.doi.org/10.1016/0550-3213(96)00369-0}{{\em
  Nucl. Phys.} {\bf B476} (1996)  437--469},
\href{http://arxiv.org/abs/hep-th/9603161}{{\tt arXiv:hep-th/9603161
  [hep-th]}}.

\bibitem{Denef:2008wq}
F.~Denef, ``{Les Houches Lectures on Constructing String Vacua},'' {\em Les
  Houches} {\bf 87} (2008)  483--610,
\href{http://arxiv.org/abs/0803.1194}{{\tt arXiv:0803.1194 [hep-th]}}.

\bibitem{Weigand:2018cod}
T.~Weigand, ``{F-theory},''
\href{http://dx.doi.org/10.22323/1.305.0016}{{\em PoS} {\bf TASI2017} (2018)
  016}.

\bibitem{Monitor:2017mdv}
{\bf LIGO Scientific, Virgo, Fermi-GBM, INTEGRAL} Collaboration, B.~P. Abbott
  {\em et al.}, ``{Gravitational Waves and Gamma-rays from a Binary Neutron
  Star Merger: GW170817 and GRB 170817A},''
  \href{http://dx.doi.org/10.3847/2041-8213/aa920c}{{\em Astrophys. J.} {\bf
  848} (2017) no.~2, L13},
\href{http://arxiv.org/abs/1710.05834}{{\tt arXiv:1710.05834 [astro-ph.HE]}}.

\bibitem{Nakahara:2003np}
M.~Nakahara, {\em ``Geometry, topology and physics - 2nd Edition''} (2003) Taylor and Francis group.

\bibitem{Giudice:2008bi}
G.~F. Giudice, ``{Naturally Speaking: The Naturalness Criterion and Physics at
  the LHC},''
\href{http://arxiv.org/abs/0801.2562}{{\tt arXiv:0801.2562 [hep-ph]}}.

\bibitem{Borges:libro}
J.~L. Borges, ``{La Biblioteca di Babele},''. ''Finzioni'', Einaudi, 1955.

\bibitem{Kirsten:2001wz}
K.~Kirsten, ``{Spectral functions in mathematics and physics},''. Chapman and
  Hall/CRC, Boca Raton, FL, 2001.

\bibitem{Klebanov:1998hh}
I.~R. Klebanov and E.~Witten, ``{Superconformal field theory on three-branes at
  a Calabi-Yau singularity},''
  \href{http://dx.doi.org/10.1016/S0550-3213(98)00654-3}{{\em Nucl. Phys.} {\bf
  B536} (1998)  199--218},
\href{http://arxiv.org/abs/hep-th/9807080}{{\tt arXiv:hep-th/9807080
  [hep-th]}}.

\bibitem{Scardigli:libro}
F.~Scardigli, G.~'t~Hooft, E.~Severino, and P.~Coda, ``{Determinism and Free
  Will},'' {\em Springer} (2019)  1--119.

\bibitem{Kardar:libro1}
M.~Kardar, ``{Statistical physics of particles},''. Cambridge, 2007.

\bibitem{Kardar:libro2}
M.~Kardar, ``{Statistical physics of fields},''. Cambridge, 2007.

\bibitem{Jackson:1998nia}
J.~D. Jackson, ``{Classical Electrodynamics},''. Wiley, 1998.

\bibitem{Sen:1996vd}
A.~Sen, ``{F theory and orientifolds},''
  \href{http://dx.doi.org/10.1016/0550-3213(96)00347-1}{{\em Nucl. Phys.} {\bf
  B475} (1996)  562--578},
\href{http://arxiv.org/abs/hep-th/9605150}{{\tt arXiv:hep-th/9605150
  [hep-th]}}.

\bibitem{Jizba:2014taa}
P.~Jizba, H.~Kleinert, and F.~Scardigli, ``{Inflationary cosmology from quantum
  Conformal Gravity},''
  \href{http://dx.doi.org/10.1140/epjc/s10052-015-3441-6}{{\em Eur. Phys. J.}
  {\bf C75} (2015) no.~6, 245},
\href{http://arxiv.org/abs/1410.8062}{{\tt arXiv:1410.8062 [hep-th]}}.

\bibitem{Blasone:2009ky}
M.~Blasone, P.~Jizba, F.~Scardigli, and G.~Vitiello, ``{Dissipation and
  quantization for composite systems},''
  \href{http://dx.doi.org/10.1016/j.physleta.2009.09.016}{{\em Phys. Lett.}
  {\bf A373} (2009)  4106--4112},
\href{http://arxiv.org/abs/0905.4078}{{\tt arXiv:0905.4078 [quant-ph]}}.

\bibitem{Blasone:2009yp}
M.~Blasone, P.~Jizba, and F.~Scardigli, ``{Can quantum mechanics be an emergent
  phenomenon?},'' \href{http://dx.doi.org/10.1088/1742-6596/174/1/012034}{{\em
  J. Phys. Conf. Ser.} {\bf 174} (2009)  012034},
\href{http://arxiv.org/abs/0901.3907}{{\tt arXiv:0901.3907 [quant-ph]}}.

\bibitem{Abbott:2016blz}
{\bf Virgo, LIGO Scientific} Collaboration, B.~P. Abbott {\em et al.},
  ``{Observation of Gravitational Waves from a Binary Black Hole Merger},''
  \href{http://dx.doi.org/10.1103/PhysRevLett.116.061102}{{\em Phys. Rev.
  Lett.} {\bf 116} (2016) no.~6, 061102},
\href{http://arxiv.org/abs/1602.03837}{{\tt arXiv:1602.03837 [gr-qc]}}.

\bibitem{Caswell:1929aa}
A.~Caswell {\em Science} {\bf 69} (1929)  384.

\bibitem{Albeverio:1983aa}
S.~Albeverio, P.~Blanchard, and R.~Hoegh-Krohn, ``{A stochastic model for the
  orbits of planets and satellites: an interpretation of Titius-Bode law},''
  {\em Exposition. Math.} {\bf 1(4)} (1983)  365--373.
  
\bibitem{Wilczek:2012jt} 
  F.~Wilczek,
  ``Quantum Time Crystals,''
  Phys.\ Rev.\ Lett.\  {\bf 109}, 160401 (2012)
  doi:10.1103/PhysRevLett.109.160401
  [arXiv:1202.2539 [quant-ph]].

\bibitem{Freidel:2003sp}
L.~Freidel, J.~Kowalski-Glikman, and L.~Smolin, ``{2+1 gravity and doubly
  special relativity},''
  \href{http://dx.doi.org/10.1103/PhysRevD.69.044001}{{\em Phys. Rev.} {\bf
  D69} (2004)  044001},
\href{http://arxiv.org/abs/hep-th/0307085}{{\tt arXiv:hep-th/0307085
  [hep-th]}}.

\bibitem{Nelson:1985aa}
E.~Nelson, {\em {Quantum fluctuations}}.
\newblock Princeton University Press, 1985.

\bibitem{Godel:1949ga}
K.~Godel
\href{http://dx.doi.org/10.1103/RevModPhys.21.447}{{\em Rev. Mod. Phys.} {\bf
  21} (1949)  447--450}.

\bibitem{Gama:2017eip}
F.~S. Gama, J.~R. Nascimento, A.~{\relax Yu}. Petrov, P.~J. Porfirio, and A.~F.
  Santos, ``{G\"odel-type solutions within the f(R,Q) gravity},''
\href{http://arxiv.org/abs/1707.03440}{{\tt arXiv:1707.03440 [hep-th]}}.

\bibitem{Amendola:2007qs}
L.~Amendola, M.~Gasperini, and C.~Ungarelli, ``{Non-local dilaton coupling to
  dark matter: Cosmic acceleration and pressure backreaction},''
  \href{http://dx.doi.org/10.1103/PhysRevD.77.123526}{{\em Phys. Rev.} {\bf
  D77} (2008)  123526},
\href{http://arxiv.org/abs/0711.5022}{{\tt arXiv:0711.5022 [gr-qc]}}.

\bibitem{Sornette:1997pb}
D.~Sornette, ``{Discrete scale invariance and complex dimensions},''
  \href{http://dx.doi.org/10.1016/S0370-1573(97)00076-8}{{\em Phys. Rept.} {\bf
  297} (1998)  239--270},
\href{http://arxiv.org/abs/cond-mat/9707012}{{\tt arXiv:cond-mat/9707012
  [cond-mat.stat-mech]}}.

\bibitem{Barbour:2011dn}
J.~Barbour, ``{Shape Dynamics: An Introduction},''
\href{http://arxiv.org/abs/1105.0183}{{\tt arXiv:1105.0183 [gr-qc]}}.

\bibitem{Anderson:2017jij}
E.~Anderson, ``{The Problem of Time},'' {\em Fundam. Theor. Phys.} {\bf 190}
  (2017)  pp.--.

\bibitem{Mersini-Houghton:2012mha}
L.~Mersini-Houghton and R.~Vaas, eds., {\em {The Arrows of Time}}, vol.~172.
\newblock Springer, Berlin, 2012.

\end{thebibliography}
\end{document}